\documentclass{aa} 
\usepackage{graphicx}
\usepackage{txfonts}
\usepackage{aa}
\bibstyle{aa}
\def\logg{\hbox{$\log g$}}

\def\sn{\hbox{S/N}}  
  
\def\vsin{\hbox{$v \sin i$}}  
  
\def\kms{\hbox{km\,s$^{-1}$}}  
\def\ms{\hbox{m\,s$^{-1}$}}

\def\em{\it}  
  
\def\degr{\hbox{$^\circ$}}  
\def\rpd{\hbox{rad\,d$^{-1}$}}   
   
\def\omeq{\hbox{$\Omega_{\rm eq}$}}   
\def\dom{\hbox{$d\Omega$}}   
\def\kis{\hbox{$\chi^2$}}   
\def\kisr{\hbox{$\chi^2_{\rm r}$}}   
\def\drot{\hbox{differential rotation}}   
   
\def\xib{\hbox{$\xi$~Bootis~A}}

\def\msun{\hbox{$M_\odot$}}   
\def\rsun{\hbox{$R_\odot$}}

\begin{document}
   \title{Long-term magnetic field monitoring \\ of the Sun-like star $\xi$ Bootis A \thanks{Based on observations obtained at the Bernard Lyot Telescope (TBL, Pic du Midi, France) of the Midi-Pyr\'en\'ees Observatory, which is operated by the Institut National des Sciences de l'Univers of the Centre National de la Recherche Scientifique of France.}}
   \titlerunning{Magnetic field monitoring of $\xi$ Bootis A}

   \author{A. Morgenthaler
          \inst{1,2}
          \and
          P. Petit
          \inst{1,2}  
          \and
          S. Saar
          \inst{3}
         \and
          S.K. Solanki
          \inst{4}
          \and
          J. Morin
          \inst{5,6}
          \and
          S.C. Marsden
          \inst{7}
          \and
          M. Auri\`ere
          \inst{1,2}
          \and
          B. Dintrans
          \inst{1,2}
          \and
          R. Fares
          \inst{8}
          \and
          T. Gastine
          \inst{4}
          \and  
          J. Lanoux
          \inst{1,2} 
          \and
          F. Ligni\`eres
          \inst{1,2}
          \and
          F. Paletou
          \inst{1,2}
          \and
          J.C. Ram\'irez V\'elez
          \inst{9}
          \and
          S. Th\'eado
          \inst{1,2}
          \and
          V. Van Grootel
          \inst{10}
          }

   \offprints{A. Morgenthaler}

   \institute{
   Universit\'e de Toulouse, UPS-OMP, Institut de Recheche en Astrophysique et Plan\'etologie, Toulouse, France \\ 
   \email{amorgent@ast.obs-mip.fr, petit@ast.obs-mip.fr}
\and
      CNRS, Institut de Recherche en Astrophysique et Plan\'etologie, 14 Avenue Edouard Belin, F-31400 Toulouse, France
\and
            Harvard-Smithsonian Center for Astrophysics, 60 Garden St., Cambridge, MA 02138, USA
      \and
Max-Planck-Institut f\"ur Sonnensystemforschung, Max-Planck-Str. 2, 37191 Katlenburg-Lindau, Germany
\and
 Dublin Institute for Advanced Studies, School of Cosmic Physics, 31 Fitzwilliam Place, Dublin 2, Ireland
 \and
 Institut f\"ur Astrophysik, Georg-August-Universit\"at G\"ottingen, Friedrich-Hund-Platz 1, 37077 G\"ottingen, Germany
             \and
Centre for Astronomy, School of Engineering and Physical Sciences, James Cook University, Townsville 4811, Australia
 \and 
      SUPA, School of Physics \& Astronomy, University of St Andrews, North Haugh, St Andrews KY16 9SS, UK
            \and
             Instituto de Astronomia, Universidad Nacional Autonoma de Mexico, 04510 Coyoacan, DF, Mexico           
\and
Institut d'Astrophysique et de G\'eophysique, Universit\'e de Li\`ege, 17 All\'ee du 6 Ao\^ut, B 4000 Li\`ege, Belgium
             }

   \date{Received 23 September 2011; accepted 23 February 2012}

 
  \abstract
   {}
   {We aim to investigate the long-term temporal evolution of the magnetic field of the solar-type star \xib, both from direct magnetic field measurements and from the simultaneous estimate of indirect activity indicators.}
   {We obtained seven epochs of high-resolution, circularly-polarized spectra from the NARVAL spectropolarimeter between 2007 and 2011, for a total of 76 spectra. Using approximately 6,100 photospheric spectral lines covering the visible domain, we employed a cross-correlation procedure to compute a mean polarized line profile from each spectrum. The large-scale photospheric magnetic field of the star was then modelled by means of Zeeman-Doppler Imaging, allowing us to follow the year-to-year evolution of the reconstructed magnetic topology. Simultaneously, we monitored the width of several magnetically sensitive spectral lines, the radial velocity, the line asymmetry of intensity line profiles, and the chromospheric emission in the cores of the Ca II H and H$\alpha$ lines.}
   {During the highest observed activity states, in 2007 and 2011, the large-scale field of \xib\ is almost completely axisymmetric and is dominated by its toroidal component. The toroidal component persists with a constant polarity, containing a significant fraction of the magnetic energy of the large-scale surface field through all observing epochs. The magnetic topologies reconstructed for these activity maxima are very similar, suggesting a form of short cyclicity in the large-scale field distribution. The mean unsigned large-scale magnetic flux derived from the magnetic maps varies by a factor of about 2 between the lowest and highest observed magnetic states. The chromospheric flux is less affected and varies by a factor of 1.2. Correlated temporal evolution, due to both rotational modulation and seasonal variability, is observed between the Ca II emission, the H$\alpha$ emission and the width of magnetically sensitive lines. The rotational dependence of polarimetric magnetic measurements displays a weak correlation with other activity proxies, presumably due to the different spatial scales and centre-to-limb darkening associated with polarimetric signatures, as compared to non-polarized activity indicators. Better agreement is observed on the longer term. When measurable, the differential rotation reveals a strong latitudinal shear in excess of 0.2 \rpd.}
   {}


   \maketitle

\section{Introduction}

It is generally accepted that the regular succession of magnetic minima and maxima observed in the Sun and in many cool stars is the result of an astrophysical dynamo, triggered by the combined presence of an outer convection layer and stellar rotation \citep[e.g.][]{brandenburg05}. Using chromospheric emission as a magnetic proxy, observations of stellar cyclicity have been conducted for several decades, revealing a richness in the temporal behaviour of solar-type dwarfs \citep[e.g.][]{baliunas95, saar99, lockwood07, olah09, metcalfe10}. More recently, spectropolarimetric observations of cool dwarfs have become sufficiently accurate to enable the direct detection of magnetic fields on low-activity stars \citep[e.g.][]{petit08}, further expanding our insight into magnetic variability through the ability to monitor the long-term evolution of magnetic vectorial topologies, instead of the disc-averaged chromospheric or photospheric fluxes. 

For most targets, the time-base of spectropolarimetric observations is still restricted to a few years. This limited time span can only offer a fragmentary view of magnetic cycles, except for cycle periods much shorter than solar. A few dwarfs have already been observed to undergo at least one global polarity switch \citep[e.g.][]{petit09}, or to complete a full magnetic cycle \citep{fares09, morgenthaler11}. Following the temporal evolution of large-scale stellar magnetic topologies provides important constraints for numerical simulations of stellar dynamos, particularly now that the magnetic cycles of cool stars can be investigated through 3-D MHD simulations \citep{brown10,brown11,ghizaru10}. 

Systematic comparisons between the temporal evolution of the large-scale magnetic field and other activity proxies (photospheric or chromospheric) are still largely unexplored. Such studies are of interest because different measurable quantities related to magnetic activity can carry complementary information about the magnetic field generation in cool stellar objects, in particular through the different spatial scales to which they relate. As a step in this direction, we concentrate here on the solar-type star \xib. This active star is a main-sequence dwarf with an effective surface temperature of 5600~K and a surface gravity \logg=4.65 \citep{valenti05}. As a visual binary system, the masses of $\xi$~Bootis~A and B have been determined from astrometry to be 0.85 and 0.72 \msun\, respectively \citep{wielen62}. The very high magnetic activity of \xib\ \citep{baliunas95} has allowed early magnetic field detections \citep{robinson80} and is linked to a fast rotation period of 6.43~d \citep{toner88}. Given a low \vsin\ of 3 \kms\ \citep{gray84} and a stellar radius of about 0.8 \rsun\ \citep{petit05}, the short rotation period implies a low stellar inclination angle of about 28\degr.  

In this article, we simultaneously investigate the seasonal evolution of various activity proxies of \xib\ using (a) the large-scale surface magnetic topology, along with its short-term distortion through latitudinal differential rotation, (b) the Zeeman broadening of high Land\'e factor spectral lines, (c) the Ca II H and H$\alpha$ core emission and (d) the radial velocity of intensity profiles, together with their asymmetry. We first describe the instrumental setup and spectropolarimetric time-series used in this study and the procedure employed to extract Zeeman signatures. We then detail the reconstruction of the large-scale magnetic topology of the star at seven different epochs, followed by the extraction, from the same data sets, of a number of classic activity tracers. We finally discuss the results derived from our measurements. 

\section{Instrumental setup, data reduction, and multi-line extraction of Zeeman signatures}
\label{sect:data}

\begin{figure*}[!t]
\centering
\mbox{
\includegraphics[width=4cm]{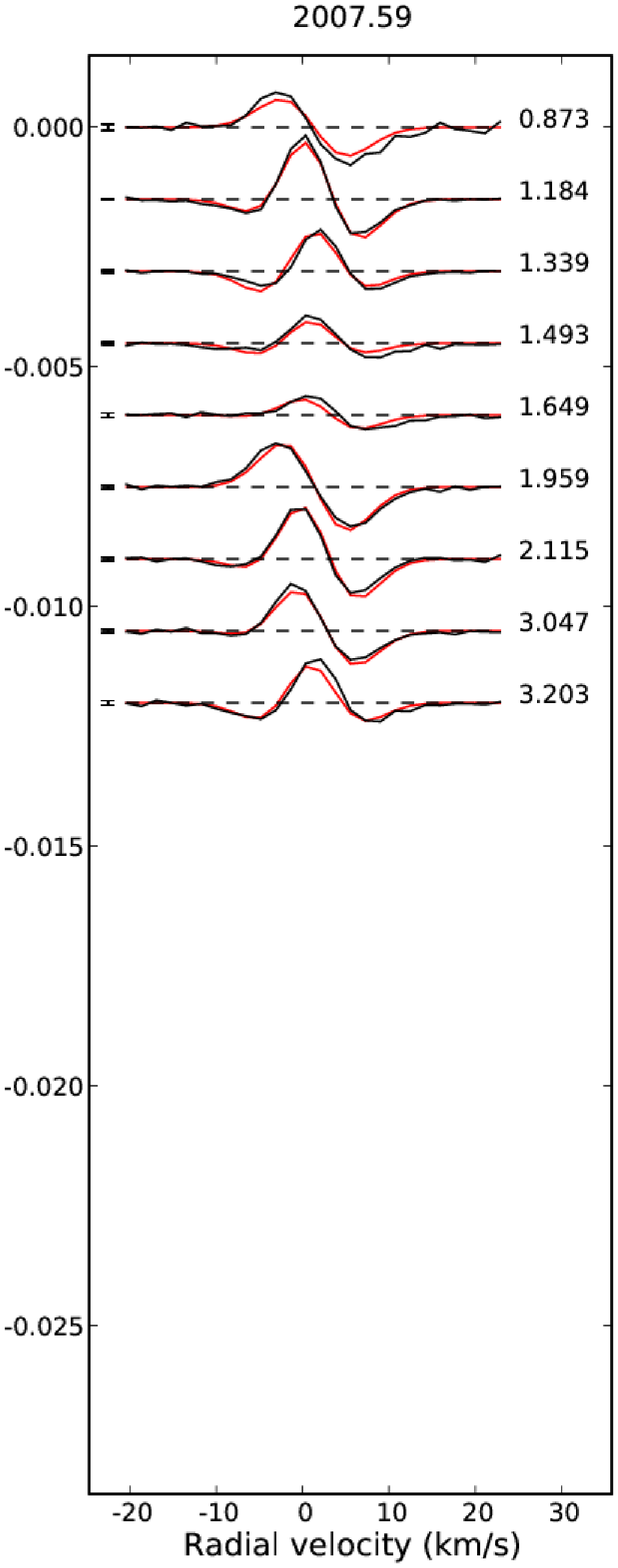}
\includegraphics[width=4cm]{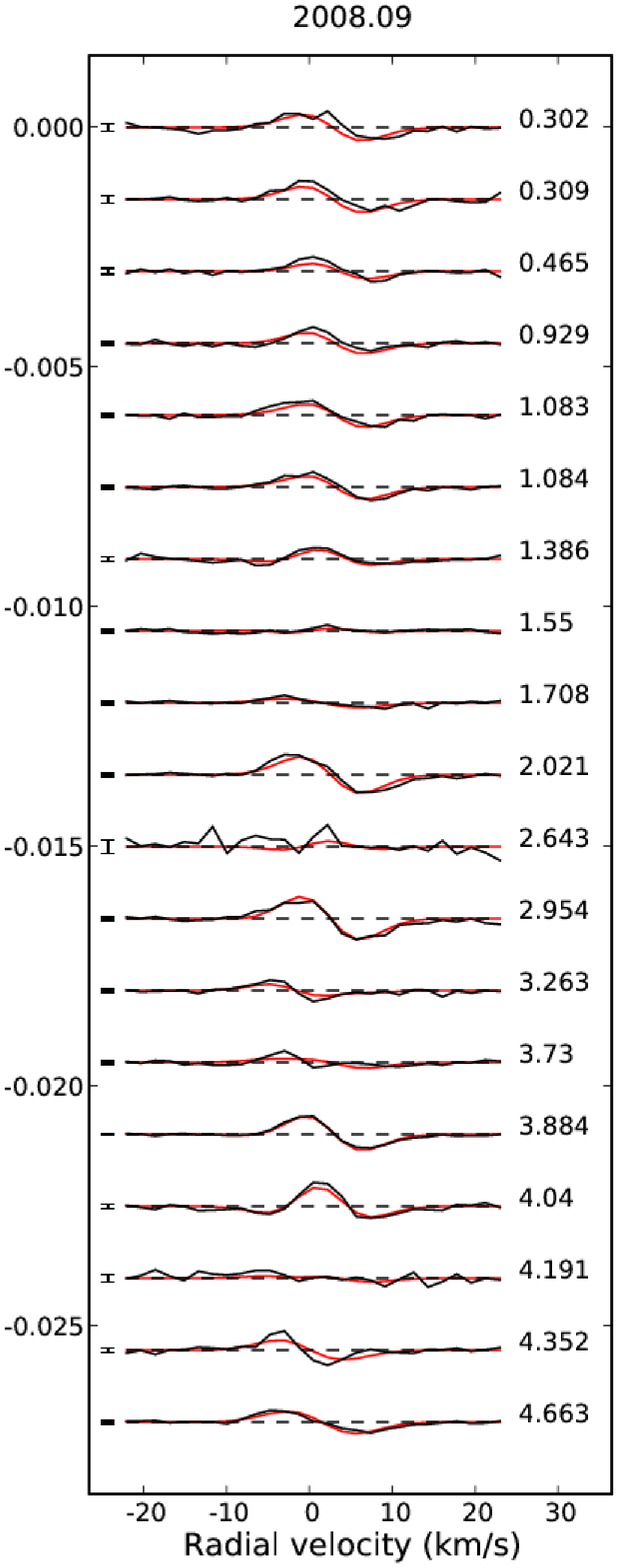}
\includegraphics[width=4cm]{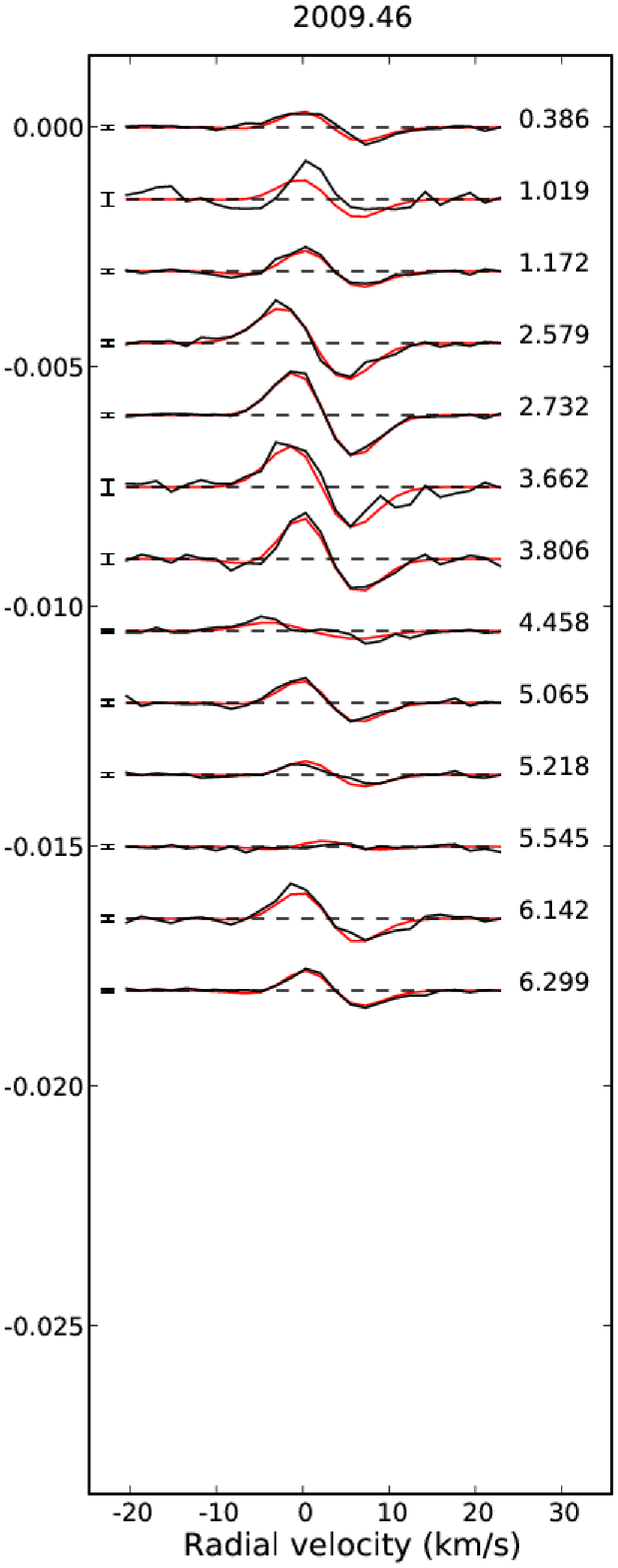}
\includegraphics[width=4cm]{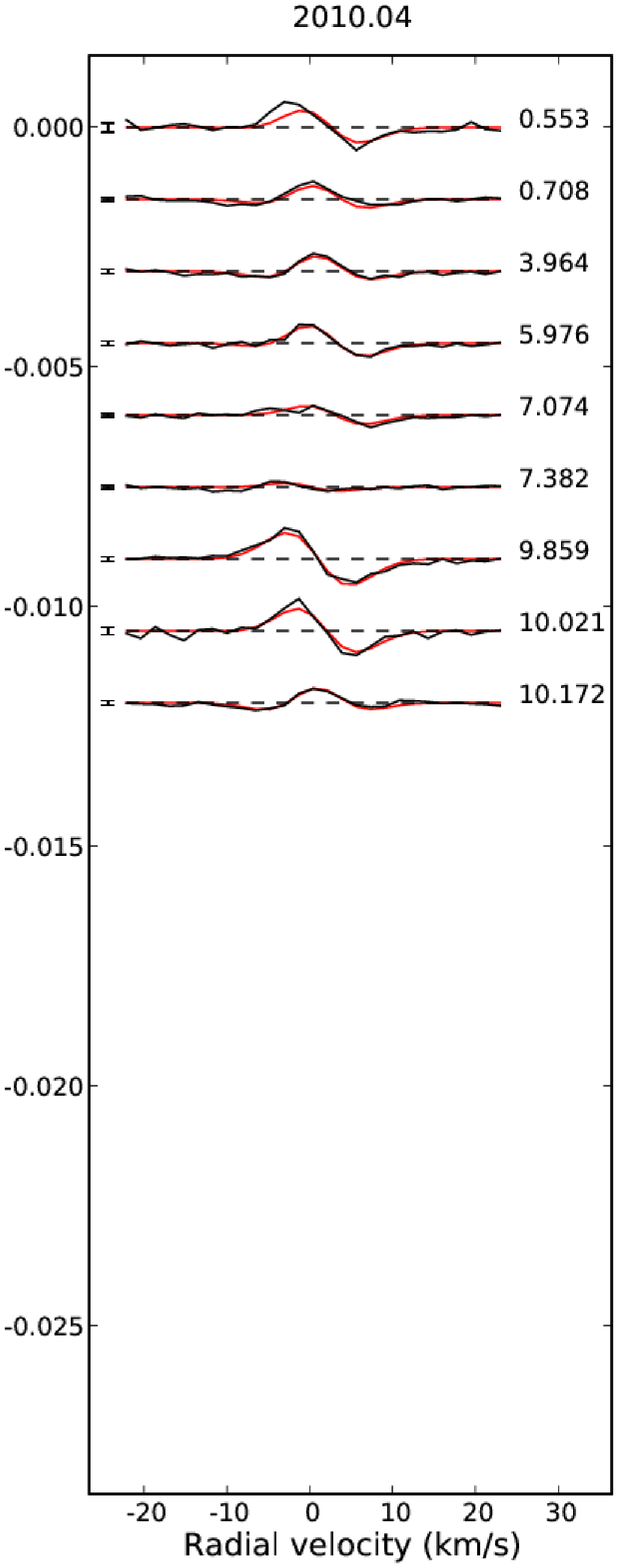}
}
\mbox{
\includegraphics[width=4cm]{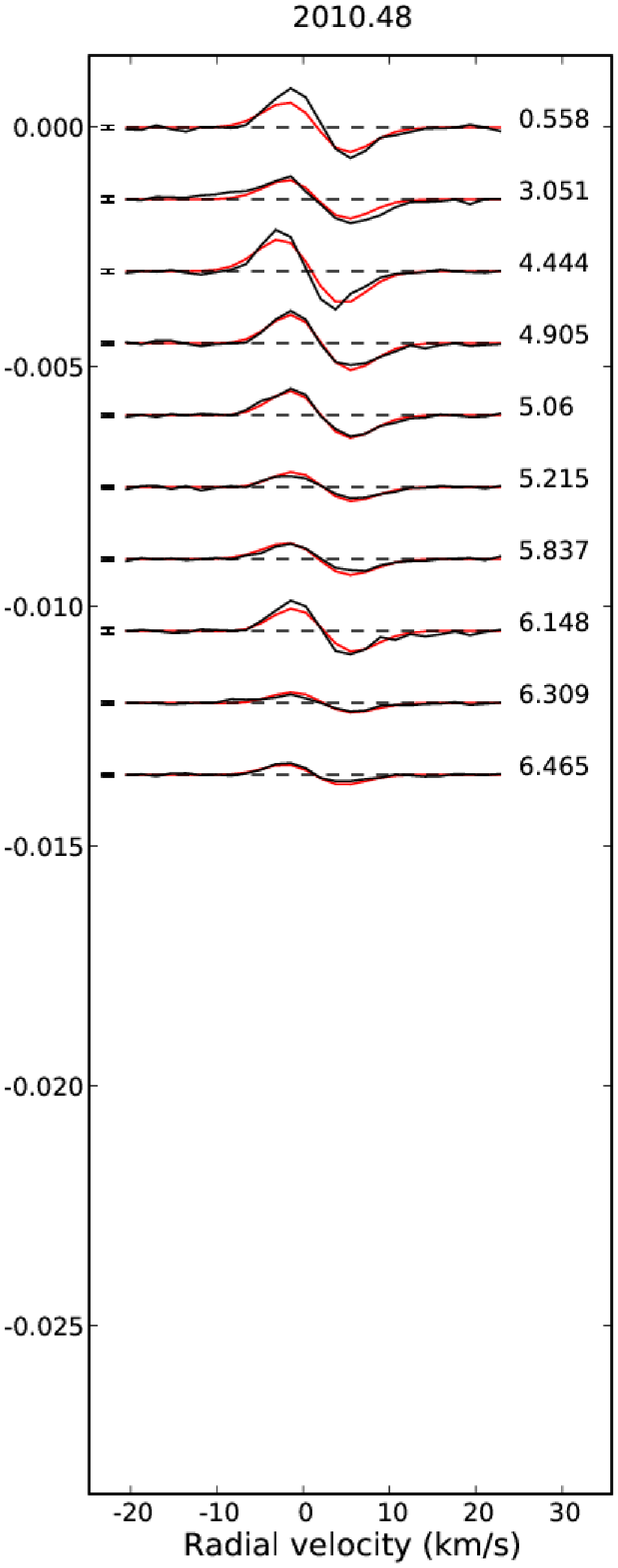}
\includegraphics[width=4cm]{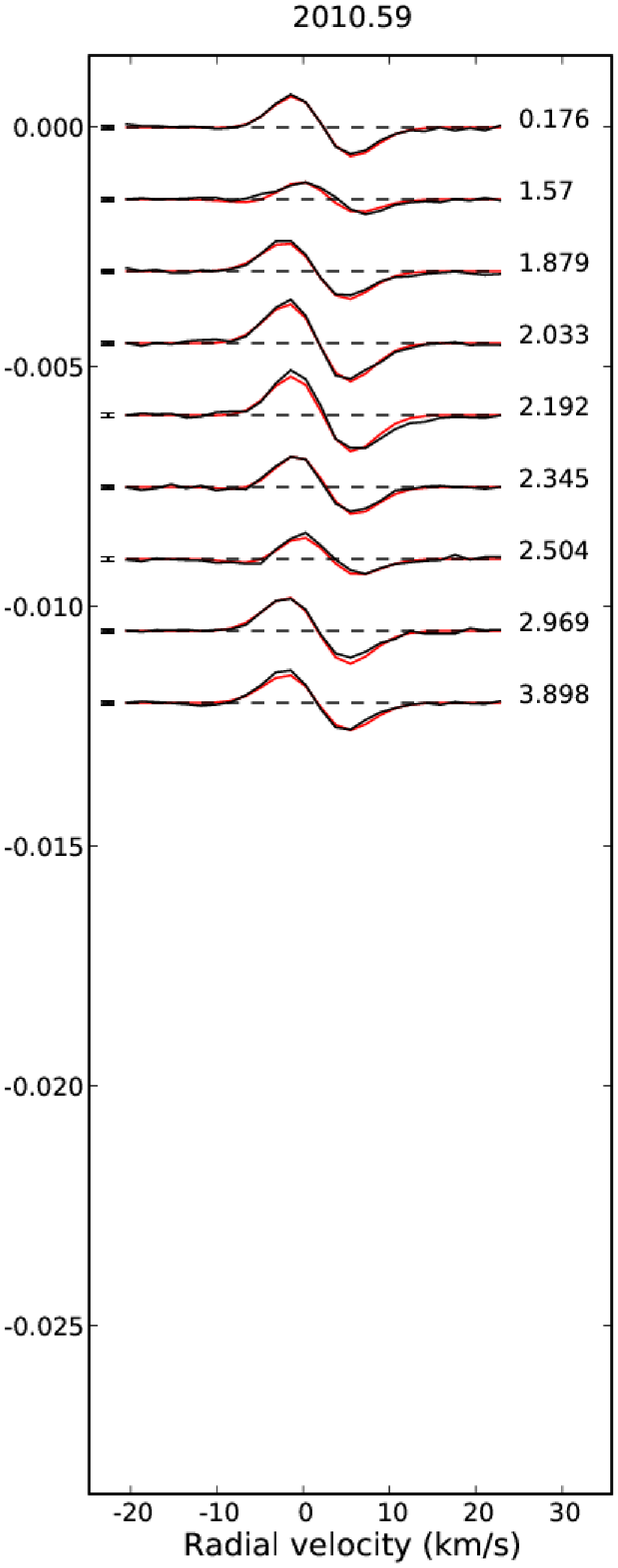}
\includegraphics[width=4cm]{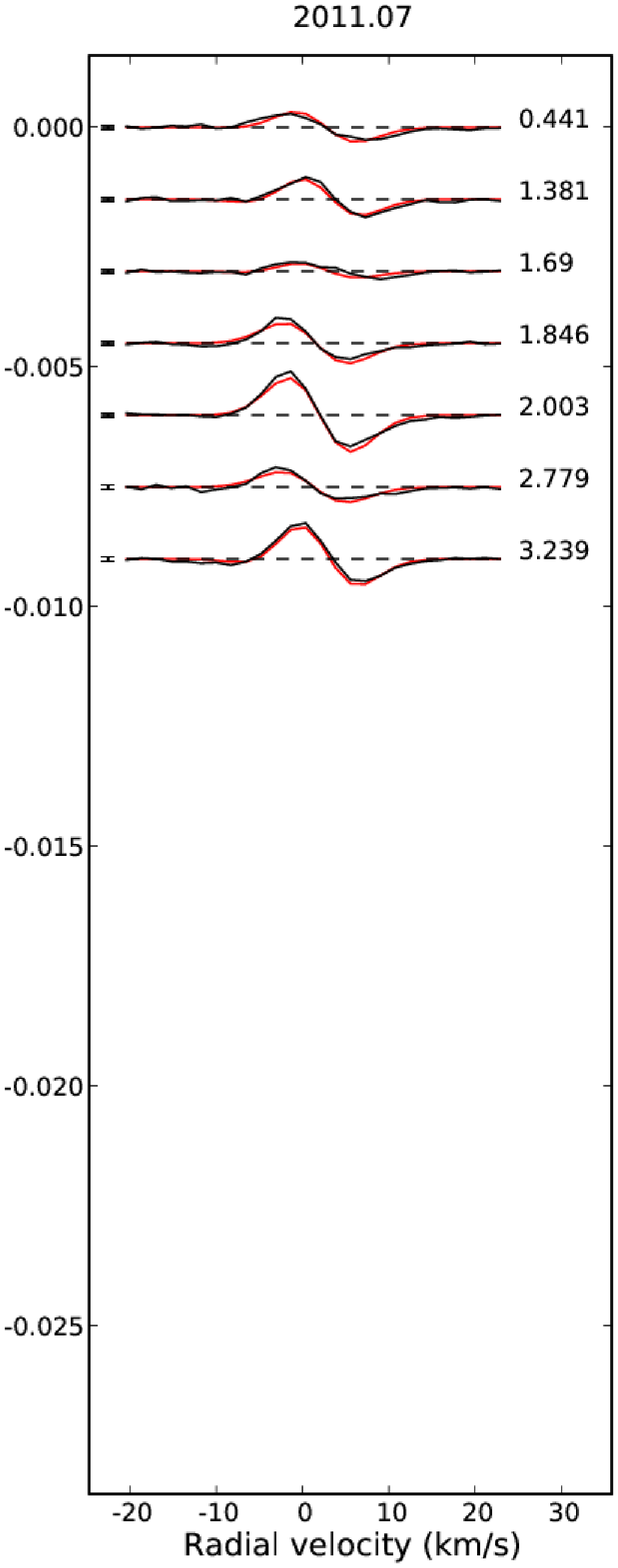}
}
\caption{Normalized Stokes V profiles ($V/I_c$) of $\xi$ Boo A for 2007.59, 2008.09, 2009.46, 2010.04, 2010.48, 2010.59 and 2011.07 (from left to right and top to bottom), after correction for the mean radial velocity of the star (the radial velocity values are listed in Tab. \ref{tab:moy_ind}). Black lines represent the data and red lines correspond to synthetic profiles of our magnetic model. Successive profiles are shifted vertically for display clarity. Rotational phases of observations are indicated in the right-hand part of the plot and error bars are illustrated on the left side of each profile.}
\label{fig:stokesv}
\end{figure*}

The data sets were collected with the NARVAL spectropolarimeter at the Telescope Bernard Lyot (Observatoire du Pic du Midi, France) whose instrumental setup is identical to the one described by \citet{petit08}. The \'echelle spectrograph has a resolution of 65,000 and covers the whole optical wavelength domain from near-ultraviolet (370 nm) to near-infrared (1,000 nm) in one exposure, with 40 orders on the CCD. NARVAL provides simultaneous recordings of the intensity spectrum (Stokes I) and one polarized spectrum (linear or circular). Here we deal with circular polarization (Stokes V parameter). Seven sets of spectra were obtained between 2007 and 2011 (see Table \ref{tab:obs1} and \ref{tab:obs2}). They each contain between 7 and 19 spectra (in 2011.07 and 2008.09, respectively), obtained over 15 to 62 consecutive days (in 2007.59 and 2010.04, respectively). 

The raw data were processed using LibreEsprit, an automatic reduction software developed for NARVAL, based on the algorithm detailed by \citet{donati97}. For the reduced spectra of \xib, the maximum signal-to-noise ratio (\sn) is achieved at a wavelength of about 730~nm. The \sn\ value depends on the adopted exposure time and weather conditions above the Pic du Midi Observatory, with typical values varying between 450 and 700. The \sn\ drops in the blue part of the spectrum, with a usual value of about 100 around the Ca II H\&K lines. The situation is better in the near-infrared, with a \sn\ close to 500 in the neighbourhood of FeI@846.8404 used later in this study.

The \sn\ of the polarized spectra is not sufficient to detect Zeeman signatures in individual spectral lines, even for photospheric lines with high Land\'e factors. To achieve a sufficiently low noise level, we used the reduced spectrum to calculate a single, cross-correlated photospheric line profile using the Least Square Deconvolution (LSD) multi-line technique \citep{donati97, kochukhov10}. We employed a line-list of about 6,100 spectral lines matching a stellar photospheric model for the spectral type of $\xi$ Bootis A (G8V). This line-list is the same as the one previously used by \citet{petit05}. Thanks to this cross-correlation approach, the noise level of the mean Stokes V line profiles is reduced by a factor of about 30 with respect to the initial spectrum, so that the resulting noise level lies in the range $3.0\times10^{-5}-1.6\times10^{-4}I_c$, where $I_{c}$ denotes the continuum intensity.

Fig. \ref{fig:stokesv} shows the seven sequences of Stokes V LSD line
profiles corresponding to the seven observing runs. The rotational phases, indicated in the right-hand part of the plot and listed in Tab. \ref{tab:obs1} and \ref{tab:obs2}, were calculated according to the ephemeris of \citet{petit05}, who adopted a rotational period of 6.43~d \citep{toner88} and a Julian date for zero rotational phase equal to 2,452,817.41. This rotation period is longer than the equatorial period derived from the modelling of the surface differential rotation (see Sect. \ref{sec:maps}), because it better represents the rotation period of the higher stellar latitudes that are expected to dominate the disc-integrated activity tracers, owing to the star's low inclination angle. The rotational modulation of most activity proxies investigated in this study is more pronounced when using the adopted period of 6.43~d.

Polarized features in the profile core, interpreted as Zeeman signatures, are detected for a large fraction of the observations, with amplitudes comfortably exceeding the noise level. It can already be seen in Fig. \ref{fig:stokesv} that the signal amplitudes are higher in 2007.59 than in the following years. If we take a closer look at the shape of the signatures, we note that the majority of them are antisymmetric about the line centre (e.g. the rotational phase 1.959 in 2007.59). A few of them are almost symmetric (phase 3.203 in 2007.59, 4.04 in 2008.09, and 10.172 in 2010.04). We note also that the sign of the signatures (symmetric or antisymmetric) is the same throughout the observations.

\begin{table*}
\caption[]{Magnetic quantities derived from the set of magnetic maps. }
\begin{tabular}{ccccccccccccc}
\hline
frac. year & Timespan & n$\phi$ & $B_{long}$ & $B_{\rm mean}$ & pol. en.        & dipole            & quad.             & oct.                  & axi.                   & \omeq                    & \dom           & \kisr        \\
    (2000+)                     &       (d) &         & (G)               & (G)                        & (\% tot)         & (\% pol)         & (\% pol)          & (\% pol)          &(\% tot)              & (\rpd)                      &  (\rpd)          &                                   \\
\hline
07.5872         &   14.99  & 10 & $8.7\pm7.1$ & $69\pm 27$          & $17\pm 3$   & $71\pm 2$   & $13\pm 1$     & $9\pm 1$    & $83\pm 3$     & -- & -- & $4.8$\\
08.0881          &    28.04  & 19 &  $4.6\pm3.1$ & $30\pm 8$          & $58\pm 6$   & $41\pm 7$   & $15\pm 1$     & $15 \pm 2$    & $56 \pm 1$     & $ 1.13 \pm 0.01 $ & $ 0.38 \pm 0.02 $ & $1.78$ \\
09.4572          &   38.02   & 13 &  $8.3\pm6.4$ & $47\pm 11$          & $35\pm 9$   & $41\pm 7$   & $20\pm 1$     & $19\pm 3$    & $69\pm 3$   & $ 1.27\pm 0.01 $ & $ 0.57\pm 0.03 $ & $1.37$\\
10.0403          &    61.85   & 9 &  $4.1\pm5.3$ & $38\pm 9$          & $32\pm 9$   & $29\pm 20$  & $9\pm 1$     & $8\pm 1$     & $29\pm 6$   & -- & -- & $1.3$\\
10.4795          &    37.98   & 10 &  $8.6\pm3.9 $ & $36\pm 12$          & $62\pm 2$   & $ 50\pm 6$  & $13\pm 3$     & $7\pm 2$     & $43\pm8 $   & $ 1.055 \pm 0.005 $ & $ 0.67 \pm 0.01$ & $2.5$\\
10.5945          &    23.93   & 9 & $11.2\pm4.2 $ & $46\pm 17$          & $13\pm 6$   & $ 48\pm 14$  & $18\pm 4$     & $13\pm 4$     & $95\pm 1$   & $ 1.09\pm 0.03$ & $ 0.27\pm 0.05 $ & $1.64$\\
11.0657          &    17.99   & 7 & $8.4\pm3.1 $ & $43\pm 20$          & $18\pm 4$   & $ 77\pm 3$  & $14\pm 1$     & $4\pm 3$     & $85\pm 2$   & -- & -- & $2.19$\\
\hline
\end{tabular}
\noindent Notes: As a function of the mean fractional year of each run, we list the timespan of successive data sets, the number of available rotation phases, the average longitudinal field (with its associated dispersion), the mean unsigned magnetic field strength ($B_{\rm mean}$), the fraction of the large-scale magnetic energy reconstructed in the poloidal field component, the fraction of the {\em poloidal} magnetic energy in the dipolar ($\ell = 1$), quadrupolar ($\ell = 2$), and octopolar ($\ell = 3$) components, and the fraction of energy stored in the axisymmetric component ($m=0$). We then list the differential rotation parameters \omeq\ and \dom\ and the \kisr\ of the magnetic models. Error bars on the ZDI magnetic quantities are derived using the method of \citet{petit08}.
\label{tab:mag}
\end{table*}

\begin{table*}
\caption[]{Mean values of the activity tracers. }
\begin{tabular}{ccccccccccc}
\hline
fractional year & $v_{\rm r}$ & Mean velocity spans & Mean line widths & $N_{CaII H}$ & log $R'_{HK}$ & $N_{H_{\alpha}}$ \\
 (2000+) & (\kms) & (\kms) & (\kms) \\
\hline
07.5872 & $1.86 \pm 0.02$ & $0.03\pm0.02$ & $15.09\pm0.28$ & $0.4434\pm0.0080$ & $-4.34\pm0.01$ & $0.3588\pm0.0020$ \\
08.0881 & $1.74 \pm 0.05$ & $0.04\pm0.02$ & $14.59\pm0.28$ & $0.4198\pm0.0079$ & $-4.38\pm0.01$ & $0.3552\pm0.0015$ \\
09.4572 & $1.88 \pm 0.03$ & $0.03\pm0.01$ & $14.34\pm0.23$ & $0.4207\pm0.0047$ & $-4.38\pm0.01$ & $0.3532\pm0.0012$ \\
10.0403 & $1.78 \pm 0.04$ & $0.03\pm0.01$ & $13.79\pm0.20$ & $0.4004\pm0.0060$ & $-4.40\pm0.01$ & $0.3513\pm0.0010$ \\
10.4795 & $1.93 \pm 0.03$ & $0.04\pm0.02$ & $14.15\pm0.16$ & $0.4032\pm0.0059$ & $-4.40\pm0.01$ & $0.3512\pm0.0011$ \\
10.5945 & $1.94 \pm 0.02$ & $0.05\pm0.01$ & $13.96\pm0.28$ & $0.4027\pm0.0114$ & $-4.40\pm0.02$ & $0.3503\pm0.0016$ \\
11.0657 & $1.83 \pm 0.02$ & $0.03\pm0.01$ & $14.52\pm0.33$ & $0.4281\pm0.0108$ & $-4.37\pm0.01$ & $0.3573\pm0.0025$ \\
\hline
\end{tabular}\\
\noindent Notes: we list the radial velocity of the star, the bisector span of Stokes I mean profiles, the width of the FeI@846.8404 line, the Ca II emission index (along with its $\log(R'_{HK})$ counterpart), and the $H{\alpha}$ emission index. All average quantities are given with their dispersion.
\label{tab:moy_ind}
\end{table*}

\section{Photospheric magnetic field}

\begin{figure*}[!t]
\centering
\mbox{
\includegraphics[width=6cm]{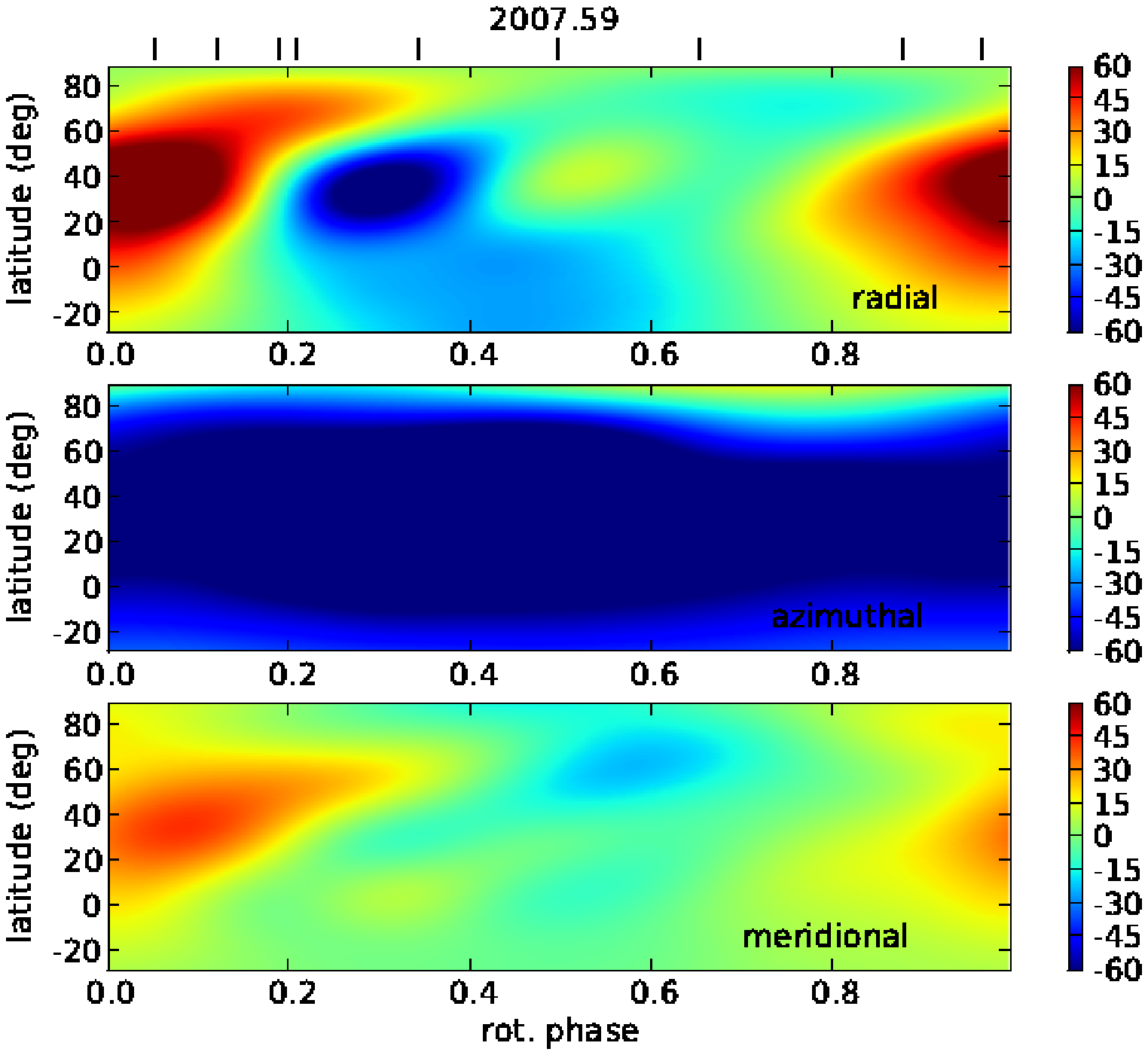}
\includegraphics[width=6cm]{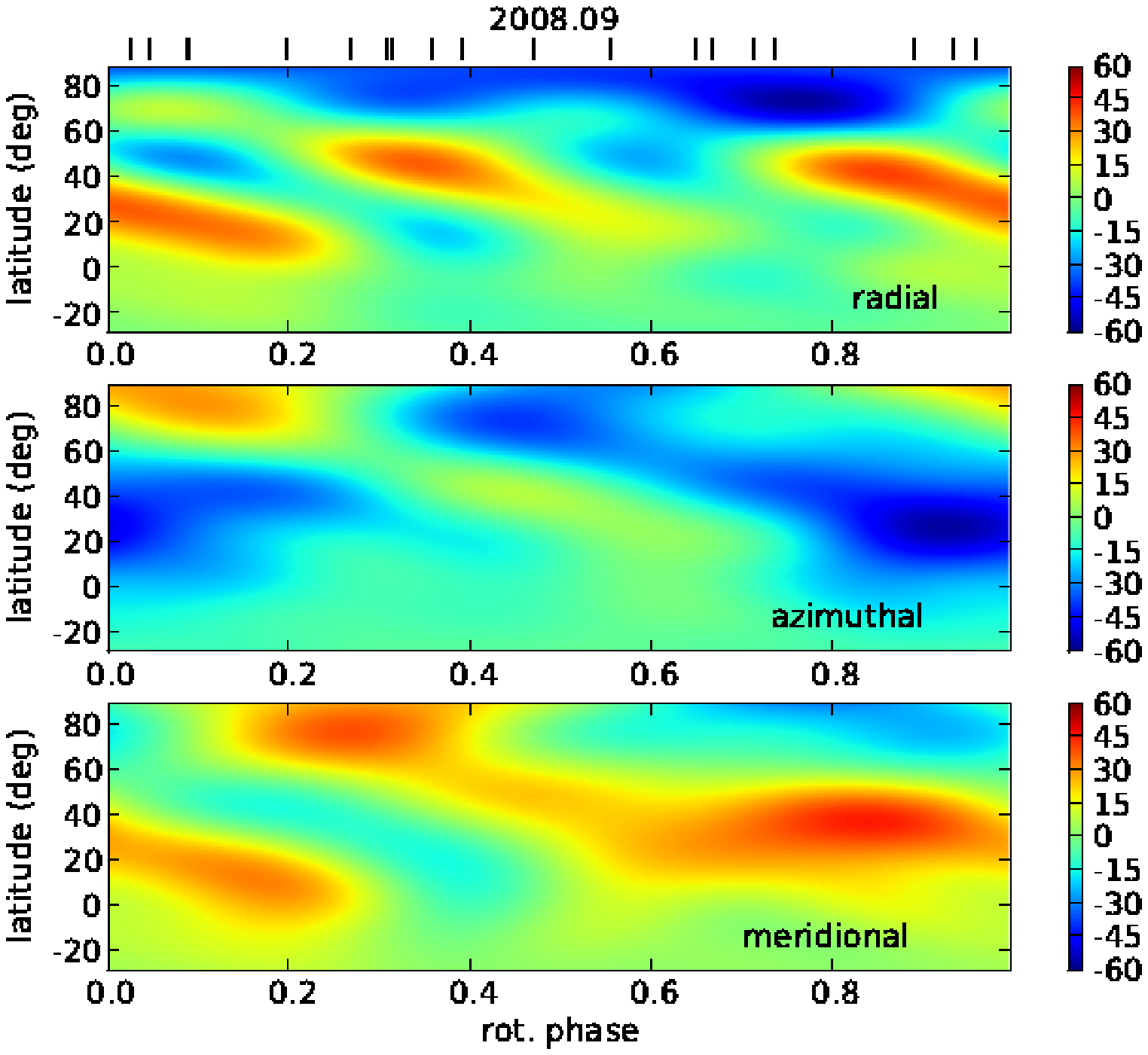}
\includegraphics[width=6cm]{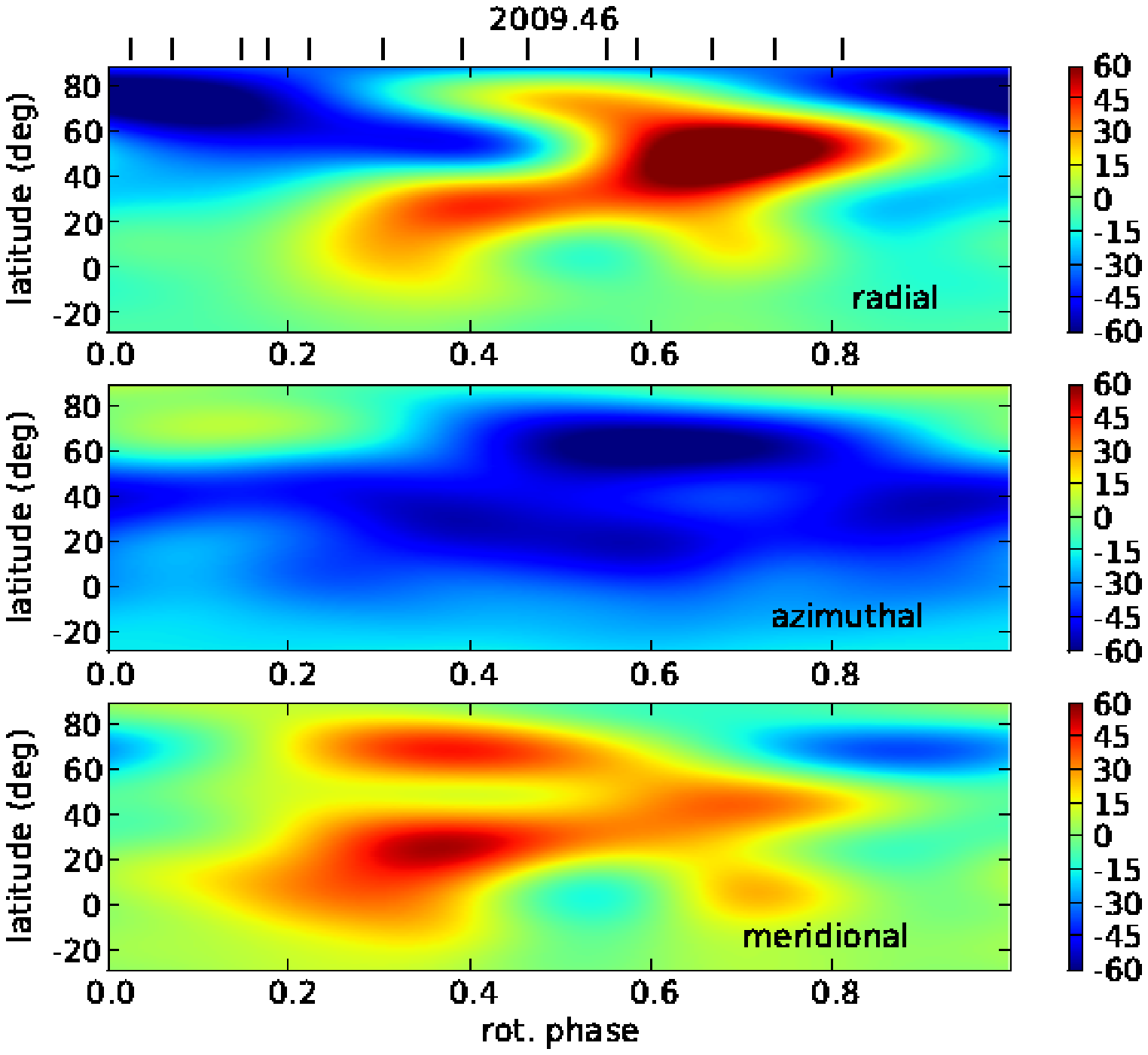}
}
\mbox{
\includegraphics[width=6cm]{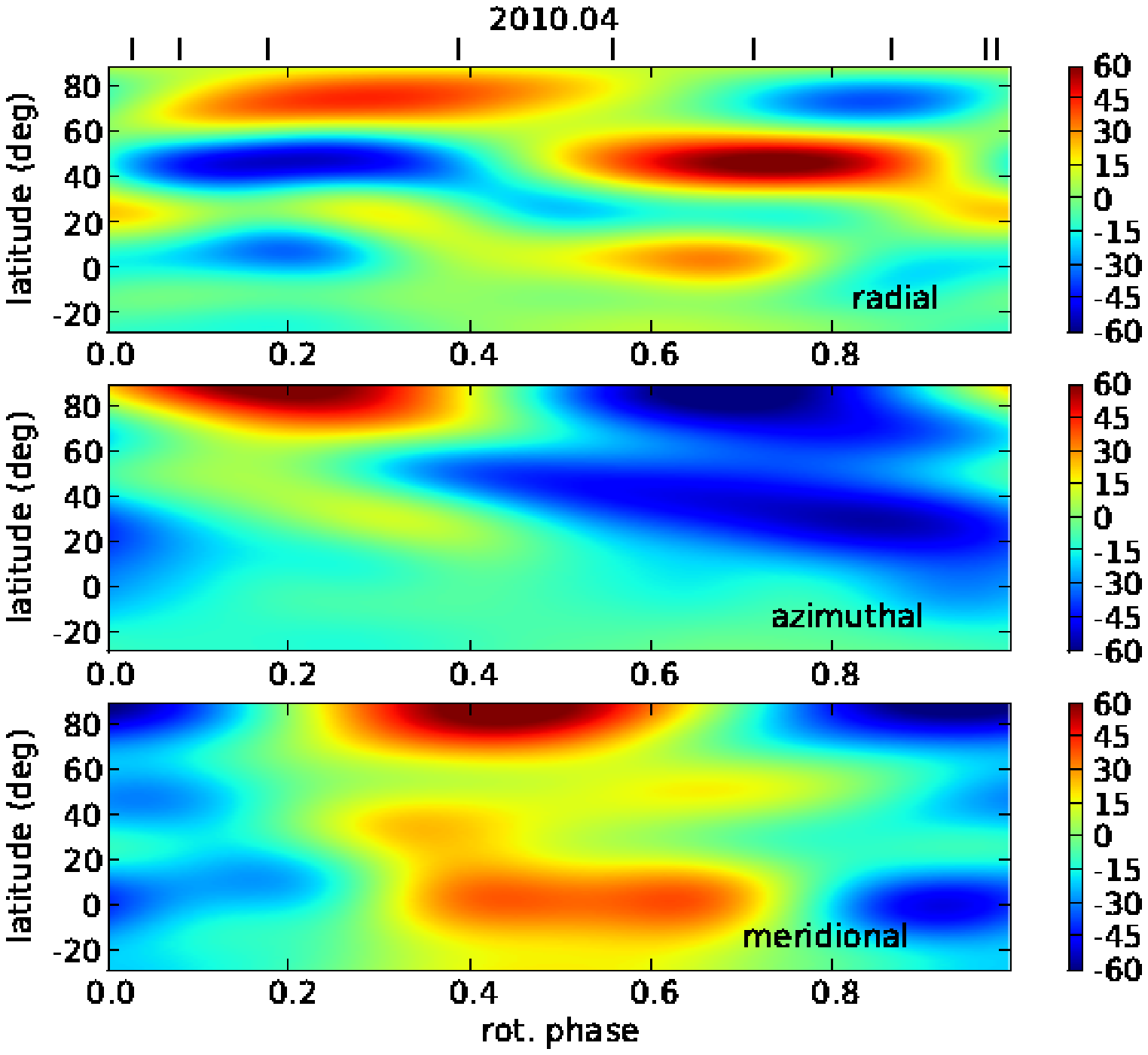}
\includegraphics[width=6cm]{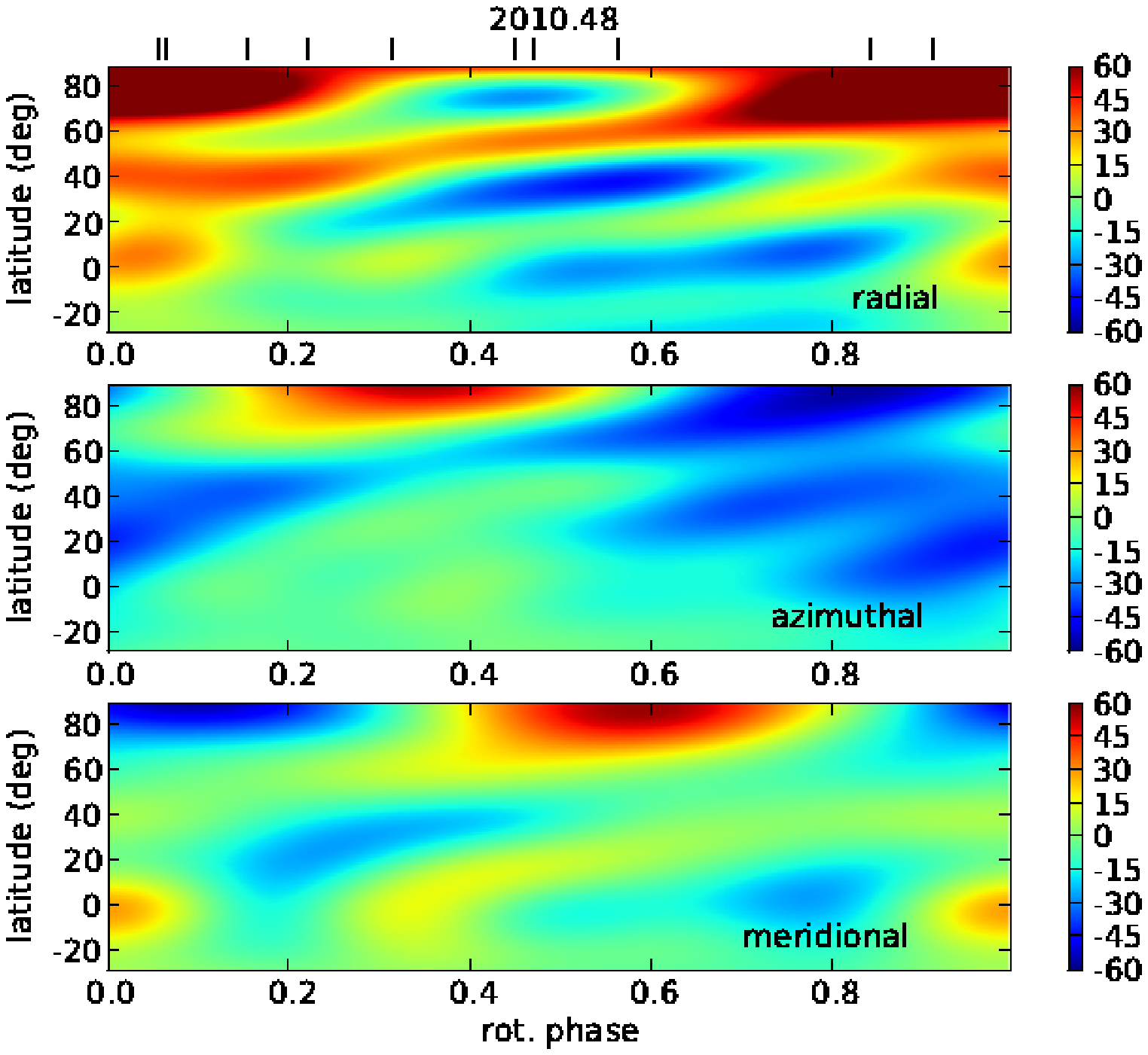}
}
\mbox{
\includegraphics[width=6cm]{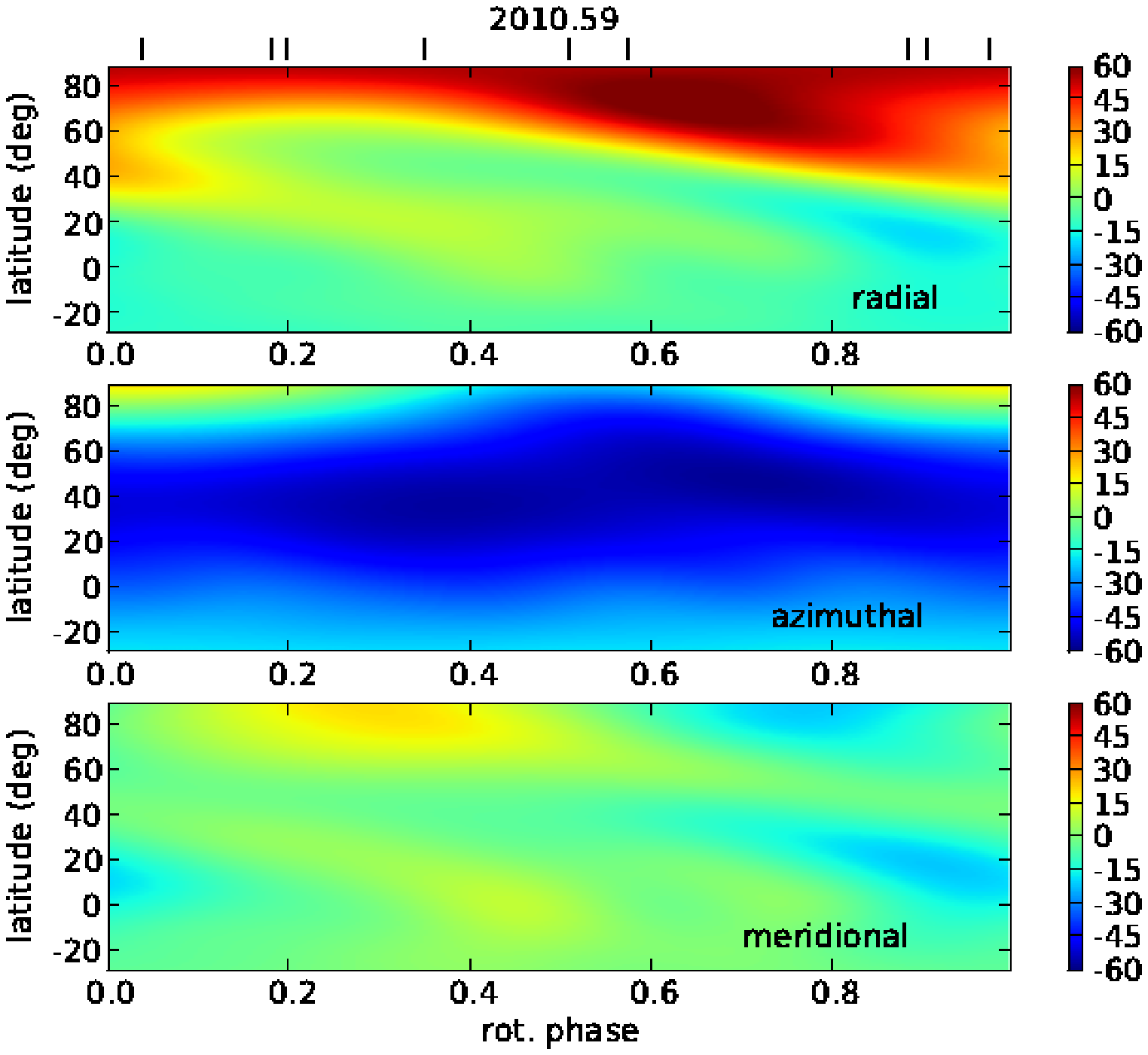}
\includegraphics[width=6cm]{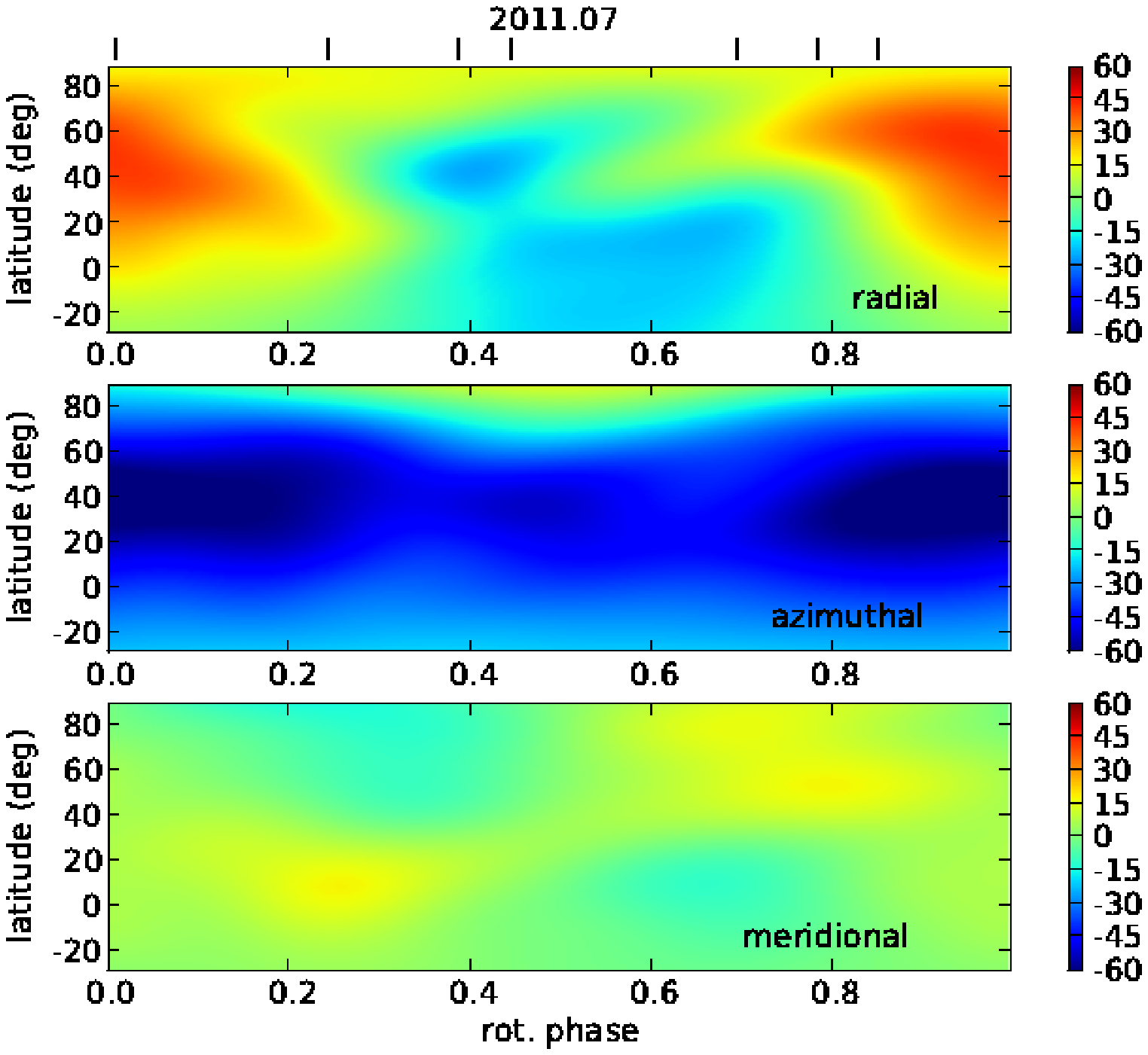}
}
\caption{Magnetic maps of $\xi$ Bootis A, derived from 2007.59, 2008.09, 2009.46, 2010.04, 2010.48, 2010.59, and 2011.07 observations (from left to right and top to bottom). For each data set, the three charts illustrate the field projection onto one axis of the spherical coordinate frame with, from top to bottom, the radial, azimuthal, and meridional field components. The magnetic field strength is expressed in Gauss.}
\label{fig:map}
\end{figure*}

\begin{figure*}[!t]
\centering
\includegraphics[scale = 0.4]{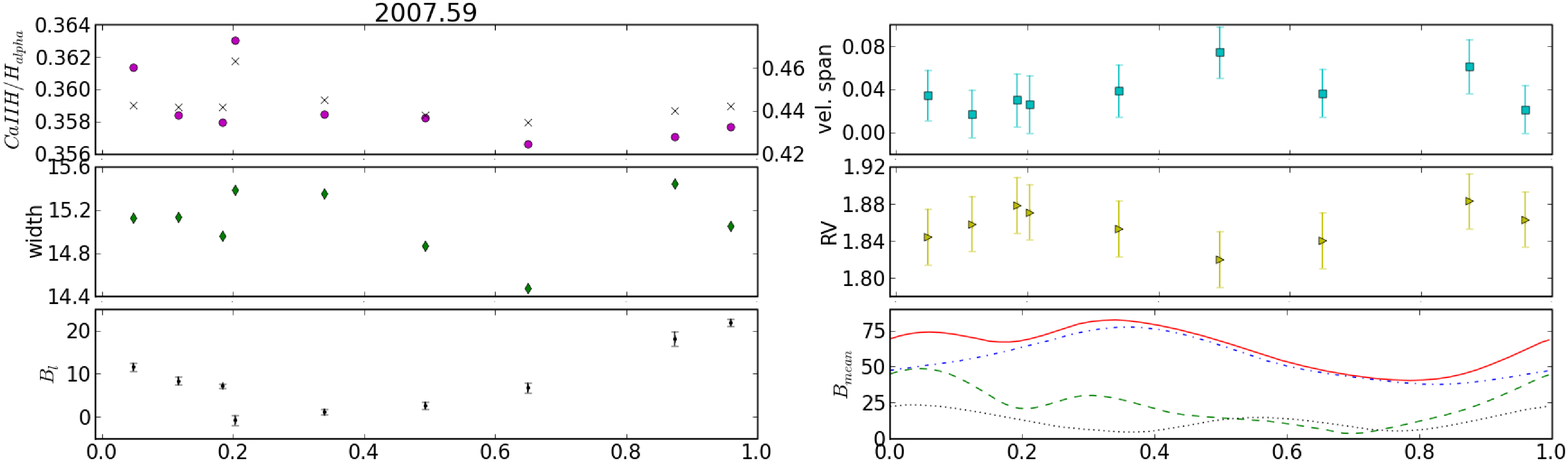}
\includegraphics[scale = 0.4]{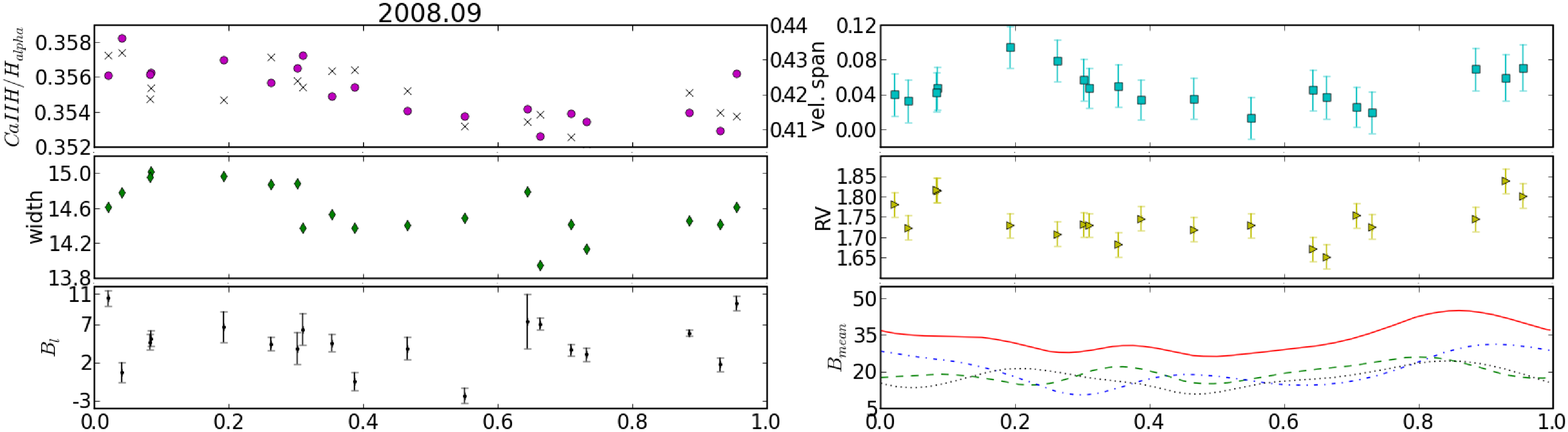}
\includegraphics[scale = 0.4]{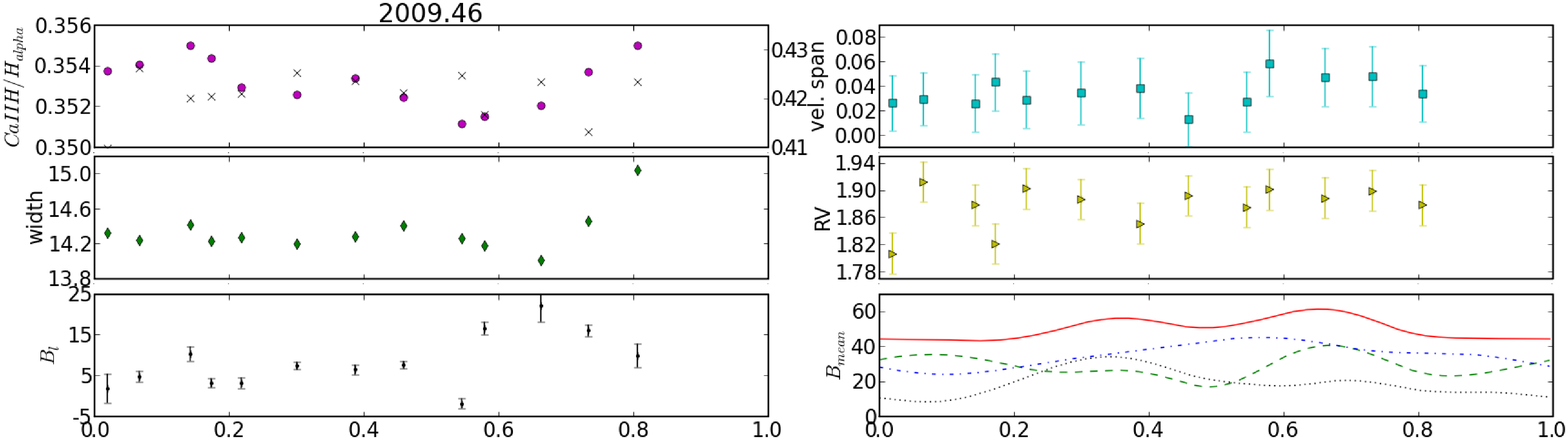}
\includegraphics[scale = 0.4]{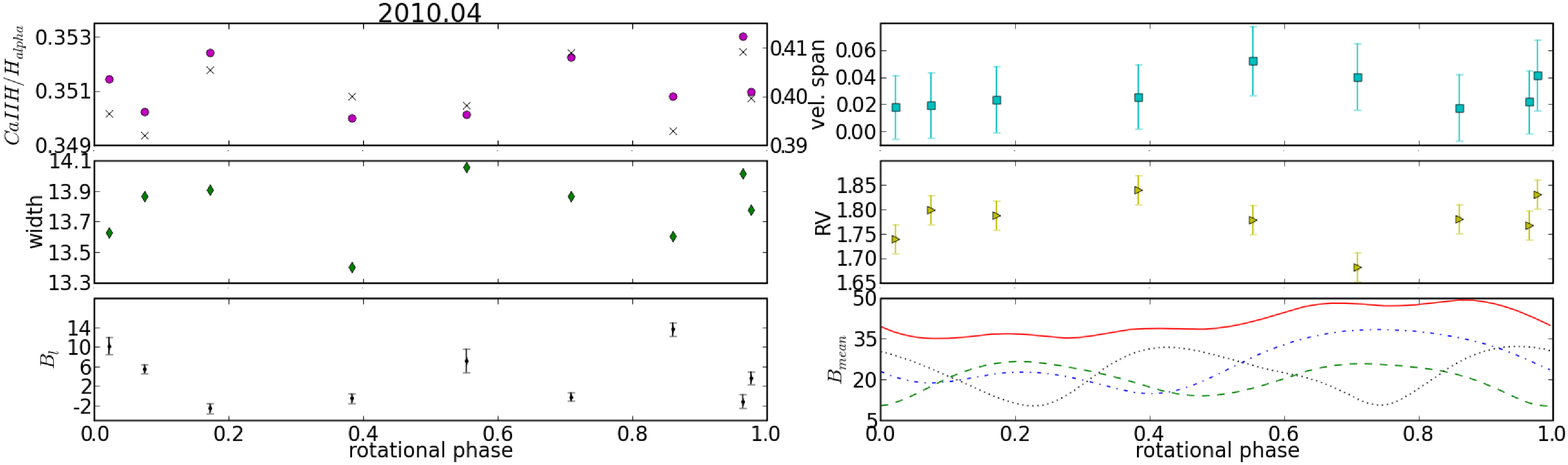}
\caption{Four sets of six subplots, each set
  corresponding to one epoch (from top to bottom, 2007.59,
  2008.09, 2009.46, and 2010.04). For each set, the left subplots
  contain, from top to bottom, the $N_{H\alpha}$-index (filled circles
  and left axis) and $N_{CaII H}$-index (crosses and right axis of the
  top panel of each set), the FeI@846.8404 magnetic line widths
  ($km.s^{-1}$), $B_l$ (Gauss). The right subplots, from top to
  bottom, correspond to the velocity spans ($km.s^{-1}$) obtained from
  the profile bisectors (Sect 4.2), the radial velocity ($km.s^{-1}$), and the mean unsigned magnetic strength of the total (full line), radial (dashes), azimuthal (dot-dashes) and meridional (dots) magnetic field components extract from the ZDI maps. Error bars are not included whenever they are smaller than the symbol size.}
\label{fig:traceurs1}
\end{figure*}

\begin{figure*}[!t]
\centering
\includegraphics[scale = 0.4]{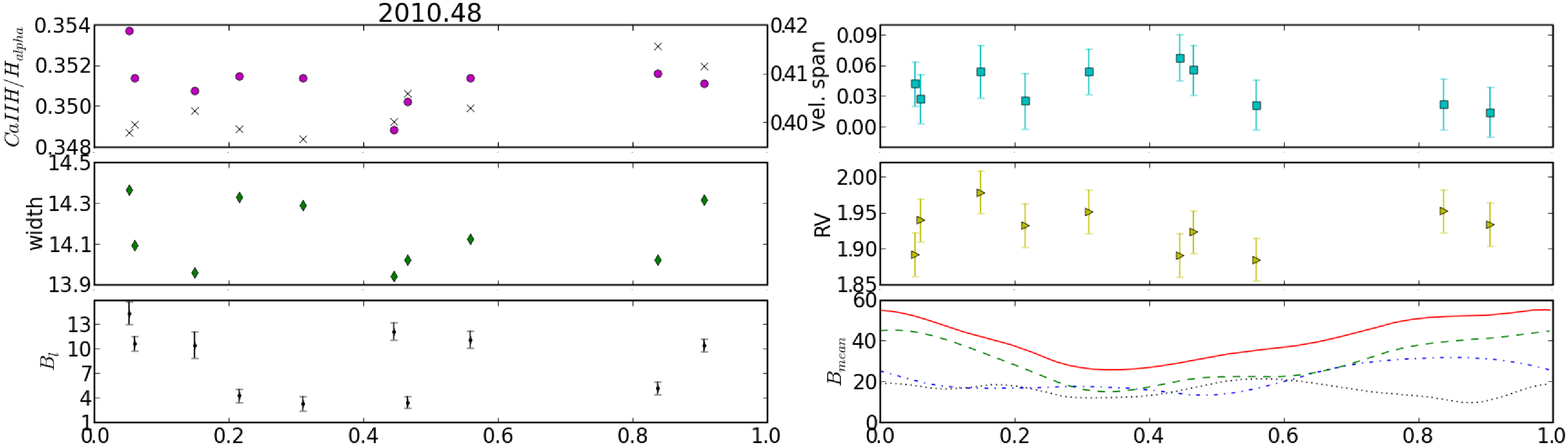}
\includegraphics[scale = 0.4]{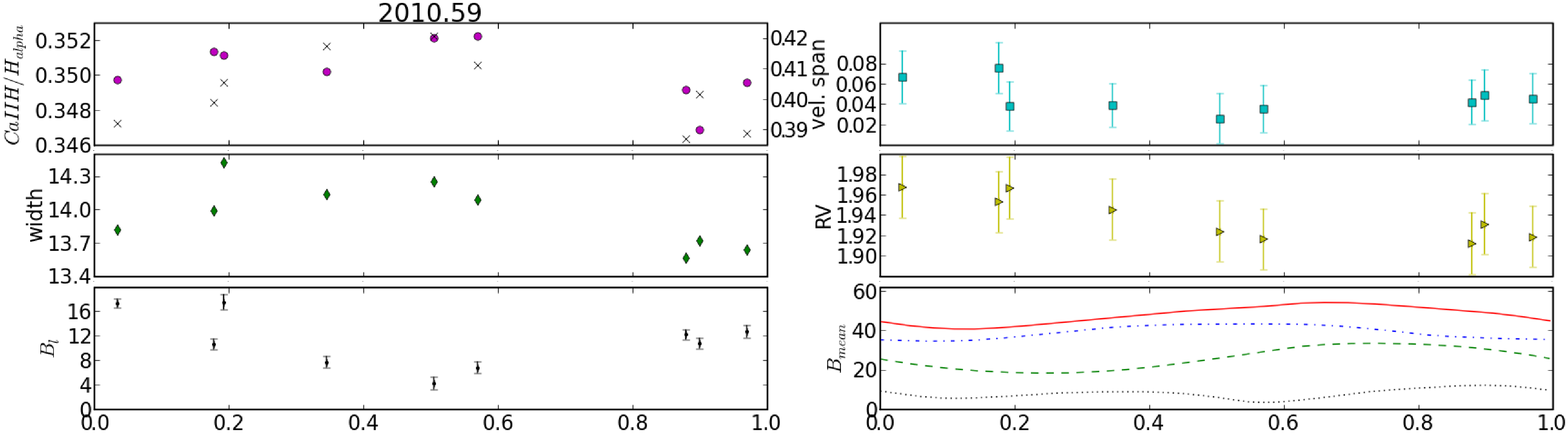}
\includegraphics[scale = 0.4]{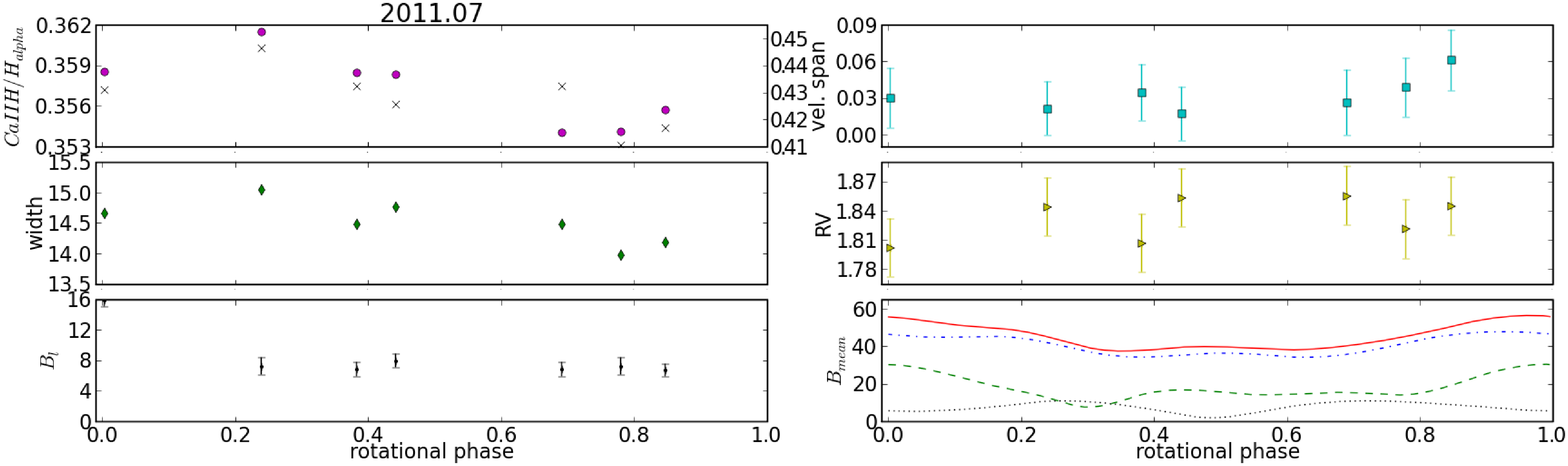}
\caption{Same as Figure \ref{fig:traceurs1} for 2010.48, 2010.59 and 2011.07.}
\label{fig:traceurs2}
\end{figure*}

\subsection{Longitudinal magnetic field}

Using the LSD line profiles in the Stokes parameters $I$ and $V$, we can derive the longitudinal component $B_l$ of the surface magnetic field, as averaged over the visible hemisphere of the star. To do so, we used the centre-of-gravity method described by \citet*{rees79}, through the equation :

\begin{equation}
B_l(G) = -2.14\times 10^{11}\frac{\int v\,V(v)\,dv}{\lambda_{0}gc\int (1-I(v))\, dv}~~,
\end{equation}

\noindent where $v$ ($km.s^{-1}$) is the radial velocity in the stellar
rest-frame, $\lambda_0$ (nm) the mean wavelength of the line mask used
to compute the LSD profile (538~nm in our case), $g$ the mean
effective Land\'e factor of the line list (equal to 1.21), and $c$
($km.s^{-1}$) the velocity of light in vacuum. The integration window covers a velocity range of $\pm 16$ \kms\ around the line centre. 

The $B_l$ values are listed in Table \ref{tab:tout} and \ref{tab:tout2}, and the averaged
values for each epoch are recorded in Table \ref{tab:mag}. We note that for Stokes V profiles with the same amplitude, longitudinal fields derived using the centre-of-gravity technique are a maximum when the Stokes V signature is anti-symmetric about the line centre. This is often not the case in our observations, because \xib\ has a large-scale toroidal field component \citep[see][and Sect. \ref{sec:maps}]{petit05}. If a profile has a higher redshift for the zero-crossing velocity, for example, phase 6.148 in 2010.48, this can result in a lower $B_l$ value, in spite of a high Stokes V amplitude.

The
rotational dependence of $B_l$ (Fig. \ref{fig:traceurs1} and
\ref{fig:traceurs2}) is visible at several epochs, and is especially
obvious in 2007.59 and 2010.59. The phase dependence is generally
more complex than a sine curve, indicating that the surface magnetic
geometry is not limited to a simple dipole. A long-term trend is also
observed in the averaged $B_l$ values (Fig. \ref{fig:val_moy}), with a
strength of 8.7 G (with a dispersion of 7.1) in 2007.59 and a much lower average value
of 4.6 G (with a dispersion of 3.1)
in 2008.09. This global decrease can readily be seen in the
decreased amplitude of the Stokes V profiles between the two
epochs. Most of the later measurements have a similar or higher $B_l$ value to that in 2007.59, with the exception of 2010.04, which has a value similar to that in 2008.09.

\subsection{Magnetic maps}
\label{sec:maps}

The rotational modulation observed in the line-of-sight projection of the magnetic field (Fig. \ref{fig:stokesv}) indicates a complex surface distribution of magnetic fields. We can model this complex field topology using a tomographic approach.

We employed Zeeman-Doppler Imaging (ZDI) to reconstruct the surface distribution of the magnetic vector from the time-series of Stokes V LSD profiles at each observing epoch. This method, first proposed by \citet*{semel89}, was implemented and tested by \citet{donatibrown97}. The version of the ZDI code used here assumes that the field geometry is projected onto a spherical harmonics frame \citep{donati06}.  In this inversion procedure, the time series of Stokes V LSD profiles is compared to a set of synthetic Stokes V line profiles computed for the same rotational phases as the observed profiles. Synthetic Stokes profiles are calculated from an artificial star whose surface is divided into a grid of pixels. Each surface pixel is associated with a local Stokes I and V profile. Assuming a given magnetic field strength and orientation for each pixel, local Stokes V profiles are calculated under the weak-field assumption, where Stokes V is proportional to $g.\lambda_0^2 .B_\parallel .\partial I / \partial \lambda$, with $\lambda_0$ representing the average wavelength of the LSD profile, $B_\parallel$ the line-of-sight projection of the local magnetic field vector and $g$ the effective Land\'e factor of the LSD profile.

We furthermore assumed that there are no large-scale brightness inhomogeneities over the stellar surface, so that all synthetic Stokes I profiles are locally the same everywhere. The limitation of this assumption will be illustrated by the observed variations in radial velocity and profile bisectors  (Sect. \ref{sec:radial}), providing us with evidence that Stokes I profiles do vary as a function of the rotational phase, due to starspots or plages. The restriction of the reconstructed magnetic topology to the global-scale component of the surface magnetic field is probably limiting the consequences of these model inaccuracies, which should affect mainly smaller, unresolved spatial scales. In the present model, the Stokes I profiles are assumed to possess a Gaussian shape, with a depth and width adjusted to achieve a good match between synthetic and observed Stokes I line profiles. We assumed a projected rotational velocity (\vsin) of 3 \kms\ and an inclination angle equal to 28\degr, both of these values were previously used in the forward modelling of the field topology made by \citet{petit05}. We also assumed the limb-darkening to be linear with $\mu = \cos(\theta)$, where $\theta$ is the limb angle, with a coefficient equal to 0.75. We finally restricted the spherical harmonics expansion to $\ell_{\rm max} \le 10$, since no improvement to the fits, between modelled and observed LSD profiles, is noticed for values of $\ell_{\rm max}$ greater than 5.

Because each data set was collected over several weeks, we assumed that the magnetic geometry might be distorted by latitudinal differential rotation over the timespan of the data collection. We therefore included a two-parameter differential rotation law in our stellar model, with the form 

\begin{equation}
\Omega(l) = \omeq - \dom \sin^2 l~~,
\end{equation} 

\noindent where $\Omega(l)$ is the rotation rate at latitude $l$, \omeq\ the rotation rate of the equator and \dom\ the difference in rotation rate between the poles and equator. Following the method of \citet{petit02}, a grid of magnetic inversions was calculated for a range of values of the quantities \omeq\ and \dom. The values listed in Tab. \ref{tab:mag} correspond to the \kis\ minimum in the parameter plane, whenever this minimum exists and is unique in the scanned \omeq-\dom\ area. 

One first limitation of this indirect imaging procedure is the roughness of the underlying stellar model, which, combined with the sparse phase sampling and uneven \sn\ ratio, may be the source of inaccuracies in the reconstructed magnetic geometry \citep[see, e.g.,][]{donatibrown97}. One other limitation of the maximum-entropy algorithm is the absence of error bars on the resulting maps. To limit the consequences of these two restrictions as much as possible, we do not discuss here the finest details of the magnetic topology (i.e., individual magnetic spots), but rather concentrate on a set of quantities derived from the largest spatial scales of the field geometry (e.g., its low-order multipolar expansion), as listed in Tab. \ref{tab:mag}. As an attempt to evaluate the uncertainties on these global magnetic quantities, we then reproduce the approach of \citet{petit08} and compute several magnetic maps, each of which is calculated using different values of the input parameters of ZDI (within the error bars on individual parameters). The dispersion of the resulting magnetic values are considered as error bars. As for other solar-like stars presented by \citet{petit08}, we generally note that the resulting dispersion is dominated by the uncertainty on the inclination angle of \xib.

The fits from our tomographic modelling are illustrated in Fig. \ref{fig:stokesv} and the magnetic maps are given in Fig. \ref{fig:map}. The modelled data were fitted to the observed ones with a reduced \kis\ (\kisr\ hereafter) of between 1.3 and 4.8 (see Tab. \ref{tab:mag}), showing that the fit accuracy never reached the noise level. The fit is not significantly improved by simple changes in the local line profile model, e.g. by the replacement of the Gaussian local Stokes I profile by a Lorentzian one (in an attempt to slightly improve the fitting of the profile wings), or the implementation of an ad-hoc asymmetry in local Stokes V profiles, as previously proposed by \citet{petit05}. A possible cause for the repeated mismatch may be the residual effect of blended lines \citep{kochukhov10} which could be larger than the noise level. Another likely cause is the intrinsic evolution of the magnetic topology during data collection, if this evolution is not entirely caused by to latitudinal differential rotation, or if the latitudinal shear is following a different law than the simple formula used here. Fast evolution of the magnetic geometry is observed between 2010.48 and 2010.59, but the field configuration seems to be more stable at other epochs, for instance in 2010.04, where a 62~d timespan produced a magnetic inversion with \kisr = 1.3. 

The limited lifetime of magnetic tracers is especially of concern in the measurement of surface differential rotation. In the ideal situation of a magnetic topology progressively distorted by a surface shear, the \kisr\ landscape in the \omeq-\dom\ plane takes the shape of a 2-D paraboloid \citep{petit02}. This is never the case here, except for 2008.09. At other epochs, iso-\kisr\ contours are sometimes not ellipsoidal (2009.46, 2010.48, 2010.59), delimit at least two local \kisr\ minima of similar depth (2011.07), or display no minimum (2007.59 and 2010.04). Therefore the apparent temporal variations of the shear shown in Table \ref{tab:mag} should be taken with caution, even when the observed differences are above the statistical error bars. In particular, we note that modifying the number of profiles used to model the surface shear (for instance, by removing the first or last profile of a time-series) often has a significant impact on the measured \dom, suggesting that intrinsic magnetic variability is affecting some of our differential rotation measurements. Whenever a unique \kisr\ minimum was not identified, the \drot\ parameters of 2008.09 were adopted in the inversion process.

The magnetic maps show the reconstructed geometry of the large-scale photospheric field. Over the years, the most recognizable feature is the azimuthal field component, organized in a fragmented or complete ring encircling the rotation axis at a latitude of about 40\degr, showing a large-scale toroidal magnetic component at
the stellar photosphere. The toroidal component contains most of the magnetic energy, except in 2008.09 and 2010.48 (Table \ref{tab:mag}), and its dominant polarity is constant for all data sets. The poloidal magnetic component is dominated by a dipole, with the dipole tending to host a larger fraction of the magnetic energy when the toroidal field component is strongest, i.e., in 2007.59 and 2011.07. The fraction of the axisymmetric field (fraction of magnetic energy stored in spherical harmonics modes with $\ell = 0$) is also evolving in correlation with the strength of the toroidal field.   

\subsection{Zeeman broadening}

To investigate the photospheric magnetic field of \xib in even greater, we looked at the broadening of several spectral lines with high magnetic sensitivity. This approach, successfully applied to \xib\ in the past \citep{robinson80}, is complementary to the Zeeman-Doppler Imaging strategy, because Zeeman broadening is sensitive to the total magnetic flux, while the polarized Zeeman signatures carry selective information about the large-scale component of the magnetic field (owing to signal cancellation for close-by regions of opposite magnetic polarity). Here, we monitored the temporal fluctuations of the Zeeman broadening. To achieve this goal, we adopted a very rough tracer of the magnetic broadening by simply following the time evolution of the line width. To do so,  we first used a cubic spline algorithm to modify the wavelength sampling of the line to obtain a grid of spectral bins offering points of equal intensity on both wings of the line profile. We then evaluated the width at various depths in the line, and finally retained the depth (0.33$I_c$ above the normalized intensity of the line center) at which the rotational modulation is the most obvious. The measured width is therefore not a standard FWHM. 

We have mainly focused on the FeI@846.8404 line, which offers a high Land\'e factor ($g=2.493$) and an infrared wavelength, which both enhance the Zeeman broadening effect. Other advantages of using this specific line are the relatively high \sn\ obtained in this part of the NARVAL spectra (close to 500) and the clean surrounding continuum, which is mostly free from telluric or photospheric lines, which ensures accurate continuum normalization. The line width measurements are listed in Table \ref{tab:tout} and plotted in Fig. \ref{fig:traceurs1} and \ref{fig:traceurs2}. A complex phase dependence of the line width is obvious at several epochs. A correlated evolution is sometimes observed with $B_l$ (e.g. in 2010.48), but this occasional similarity cannot be taken as a general trend, as illustrated by the anti-correlation observed only about a month later, in 2010.59. In addition to the rotational modulation in line width, a long-term trend is observed, with a significant decrease in the line width between 2007.59 and the following data sets. This variability is larger than the rotational variability observed at individual epochs (Fig. \ref{fig:val_moy}). This decrease agrees with the trend in decreasing magnetic field strength seen in the magnetic maps (Tab. \ref{tab:mag}).

The line width was also monitored using two other magnetically sensitive lines (FeI@549.75, g = 2.26 and FeI@550.68, g = 2.00). The variations described above are consistently recovered using these lines, although the correlation with chromospheric emission (see Section \ref{sec:chromosphere}) is generally worse than for the infrared line. We also studied three lines with a weak Land\'e factor (FeI@481.78, g = 0.48, FeI@592.74, g = 0.43 and FeI@840.14, g = 0.51) which, as expected, did not exhibit any significant temporal evolution in their width.

\section{Other activity tracers}

\subsection{Chromospheric activity}
\label{sec:chromosphere}

\begin{figure}[!t]
\centering
\includegraphics[scale = 0.45]{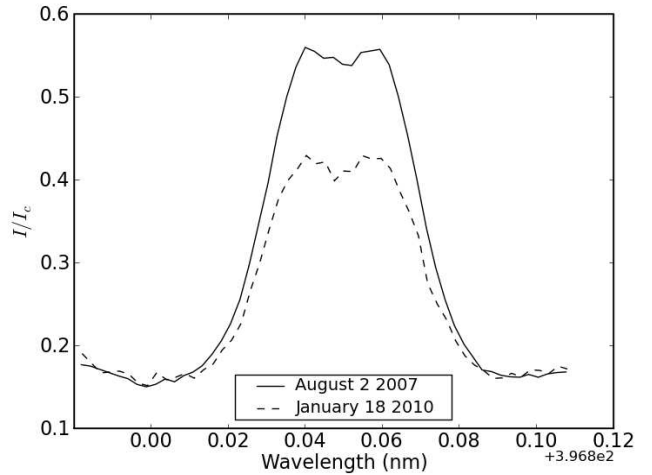}
\caption{Ca II H line for August 2 2007 (solid line) and for January 18 2010 (dashed line).}
\label{fig:Hline}
\end{figure}

Using our Stokes I spectra we monitored the evolution of the chromospheric flux between each set of observations and also across the stellar rotation cycle. A simple comparison of the core of the Ca II H line for two different years (Fig. \ref{fig:Hline}) reveals the variability of the chromospheric activity, with the core emission stronger in 2007.59 than in 2010.04. We constructed two emission indexes to quantify the emission changes, using the Ca II H and H$\alpha$ lines, respectively. Our method is described below.

\begin{figure*}[!t]
\centering
\includegraphics[scale = 0.4]{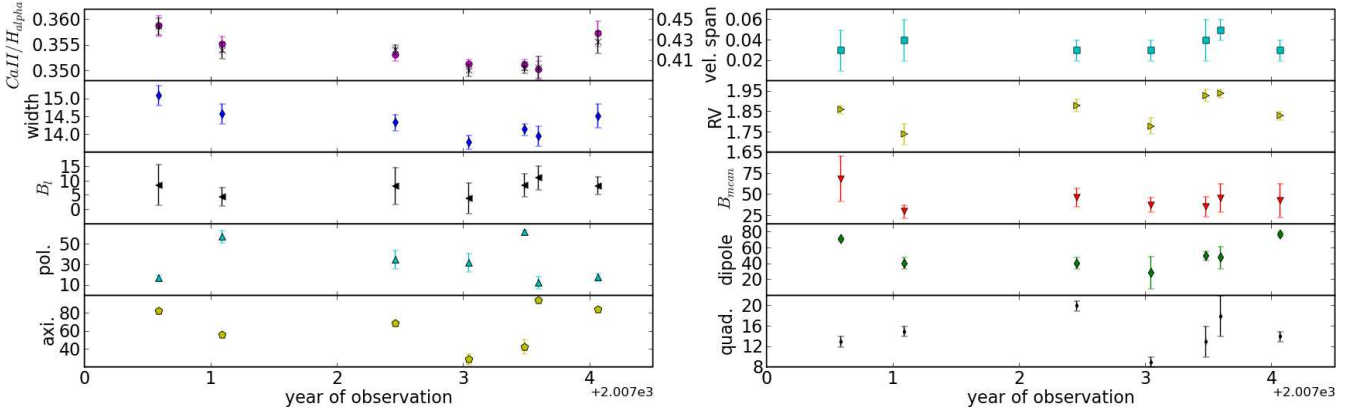}
\caption{Top three rows: long-term evolution of the average values (calculated over individual
  observing runs) and dispersion (vertical bars) of the activity proxies presented in Fig. \ref{fig:traceurs1} and \ref{fig:traceurs2}. Bottom two rows: temporal evolution of magnetic quantities listed in Tab. \ref{tab:mag}.}
\label{fig:val_moy}
\end{figure*}

\subsubsection{$N_{CaII H}$-index}

Before computing the index itself, the continuum normalization in the spectral region around Ca II H\&K was adjusted. The very dense distribution of photospheric spectral lines in this part of the spectrum prevented the standard reduction pipeline of LibreEsprit from defining a reliable continuum level, resulting in local normalization inaccuracies of the order of 20\%. To improve the situation, we took as a reference a synthetic normalized spectrum from the POLLUX database \citep{palacios10} with $T_{eff}$ and $log(g)$ values close to those of $\xi$ Boo A. First, we interpolated the synthetic spectrum on the NARVAL wavelength grid, we then defined by hand a number of reference points around the Ca II H\&K lines and used them to estimate the local ratio between the NARVAL spectrum and the synthetic spectrum. The series of ratios were then fitted by a fourth-order polynomial and, finally, the region of interest in the NARVAL spectrum was divided by the polynomial. We achieved a good continuum level using this method, without the need to rotationally broaden the synthetic spectrum. This confirms that rotational broadening can be ignored in the process, at least for a star like \xib\ that has a low \vsin. 

Afterwards, we calculated a $N_{CaII H}$-index following the method outlined in \citet{duncan91}, who defined $S$-values from Mount Wilson observations. We used two triangular bandpasses $H$ and $K$ with a FWHM of 0.1 nm to determine the flux in the line cores. Two 2 nm-wide rectangular bandpasses $R$ and $V$, centred on 400.107 and 390.107 nm, respectively, were used for the continuum flux in the red and blue sides of the H and K lines. Although results from this adapted S index are generally good, one concern was the location of Ca II K in an order overlap in the 2007.59 data set (later on, the order overlap was slighlty shifted, so that the line core fell outside the overlap region). This unfortunate position in the spectrum generates additional difficulties in obtaining a reliable continuum normalization and increases the uncertainties in the computed values (since the S/N ratio decreases rapidly towards the edge of an order). After multiple trials, we chose to use only the Ca II H line because it gave the most reliable results. Finally, in the same way as \citet{wright04} defined an $L$-index for Lick observations, we defined a $N_{CaII H}$-index for our NARVAL spectra as follows:

 \begin{equation}
N_{CaII H} = \frac{H}{R}~~,
\end{equation}

\noindent where $H$ and $R$ are the same as described above.

To match the Mount Wilson values, we transformed the $N_{CaII H}$-index as 

\begin{equation}
N_{CaII H} = \alpha\left(\frac{H}{R}\right)^2 + \beta\frac{H}{R} + \gamma~~,
\end{equation}

\noindent where $\alpha$, $\beta$ and $\gamma$ are relative weights to be determined. To estimate them, we chose 31 cool stars simultaneously present in the NARVAL archive and in  \citet{wright04}, and ensured that all selected NARVAL observations had an $S/N$ above about 100 around the Ca II H line. We then performed a least-squares fitting between the chromospheric activity values of the two stellar samples and found an optimal value of -0.972 for $\alpha$, 1.803 for $\beta$ and -0.051 for $\gamma$. Using the result of this calibration, some residual scatter is still observed between our measurements and Wright's values. These differences are probably due to the different dates of observation between the two studies, bearing in mind that the chromospheric emission of cool active stars display long-term fluctuations. The observed scatter, which is not associated to any systematic bias, is similar to that between Wright's estimates and older Mount Wilson measurements.

The $N_{CaII H}$-index is shown in Fig. \ref{fig:traceurs1} and \ref{fig:traceurs2}, and the mean values for each set appear in Table \ref{tab:moy_ind}. Random errors are about $10^{-3}$ for individual observations. Occasional repeated observations of $\xi$ Boo A during a single night (twice each night on 2008 Jan. 18 and Jan. 23) enabled us to obtain another estimate of uncertainties, assuming that the chromospheric activity is fairly constant over the few minutes that separate successive spectra. The typical difference in the $N_{CaII H}$-index between these close-by observations is about $4\times10^{-3}$. 

A rotational modulation in $N_{CaII H}$ is visible in most data sets, and is particularly evident in 2008.09. A longer-term evolution is also seen (Fig. \ref{fig:val_moy}), with differences between the years being larger than the fluctuations observed during a single rotation cycle. In individual data sets, a correlation between the rotational modulation of $N_{CaII H}$ and the width of FeI@846.84 is sometimes observed (e.g., 2008.09) but is much less pronounced at other epochs (e.g., 2009.46). A possible reason for this partial mismatch is the different 
centre-to-limb behaviour of the two magnetic field tracers. Stokes I for FeI@846.84 will be affected by changes in the central, unsplit $\pi$ component of the line (increasing towards the limb) as well as changes in the split  $\sigma$ components (decreasing towards  the limb). The combination of the two types of Zeeman components gives a different centre-to-limb variation than for the chromospheric lines, for which a limb brightening is expected if there are faculae/plages contributing to the Ca~II emission \citep{ortiz02}. The correlation between $N_{CaII H}$ and  FeI@846.84 is more easily seen on the longer term (Fig. \ref{fig:correlation}).

\subsubsection{$N_{H\alpha}$-index}

In the same manner as for $N_{CaII H}$, we defined a quantity to measure the variability in $H{\alpha}$. We used the same rectangular bandpasses that \citet{gizis02} defined around the $H{\alpha}$ line. The red-side continuum is taken to be between 656.62 and 656.84 nm, the blue-side continuum between 655.77 and 656.0 nm, and the line core between 656.10 and 656.46 nm. By defining the flux values in the continuum bandpasses as $C_{red}$ and $C_{blue}$ and the flux in the line as $F_{H\alpha}$, our index was constructed as follows:

\begin{equation}
N_{H\alpha} = \frac{F_{H\alpha}}{C_{red} + C_{blue}}~~,
\end{equation}

Note that the higher \sn\ around $H{\alpha}$, as well as the better continuum normalization in this spectral region, allowed us to skip the renormalization procedure adopted for Ca II H. The results obtained for each year are given in Figs. \ref{fig:traceurs1} and \ref{fig:traceurs2}, and the mean values are listed in Table \ref{tab:moy_ind}. Random errors are about $10^{-4}$ for a single observation. Using the two nights for which repeated observations are available, as we did for $N_{CaII H}$, the empirical uncertainty was measured to be around $7\times10^{-4}$. 

\begin{figure}[!t]
\includegraphics[scale = 0.45]{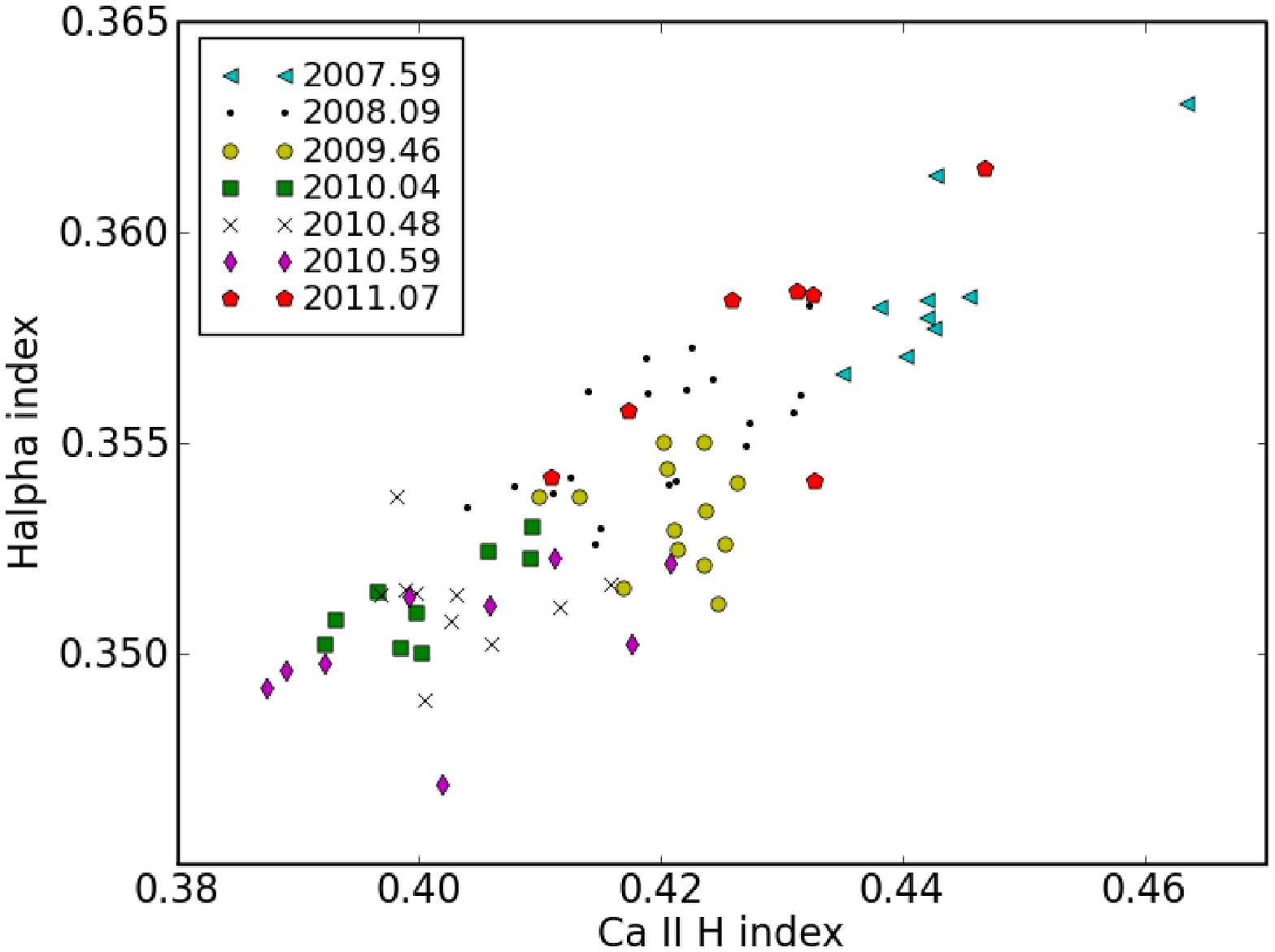}
\includegraphics[scale = 0.45]{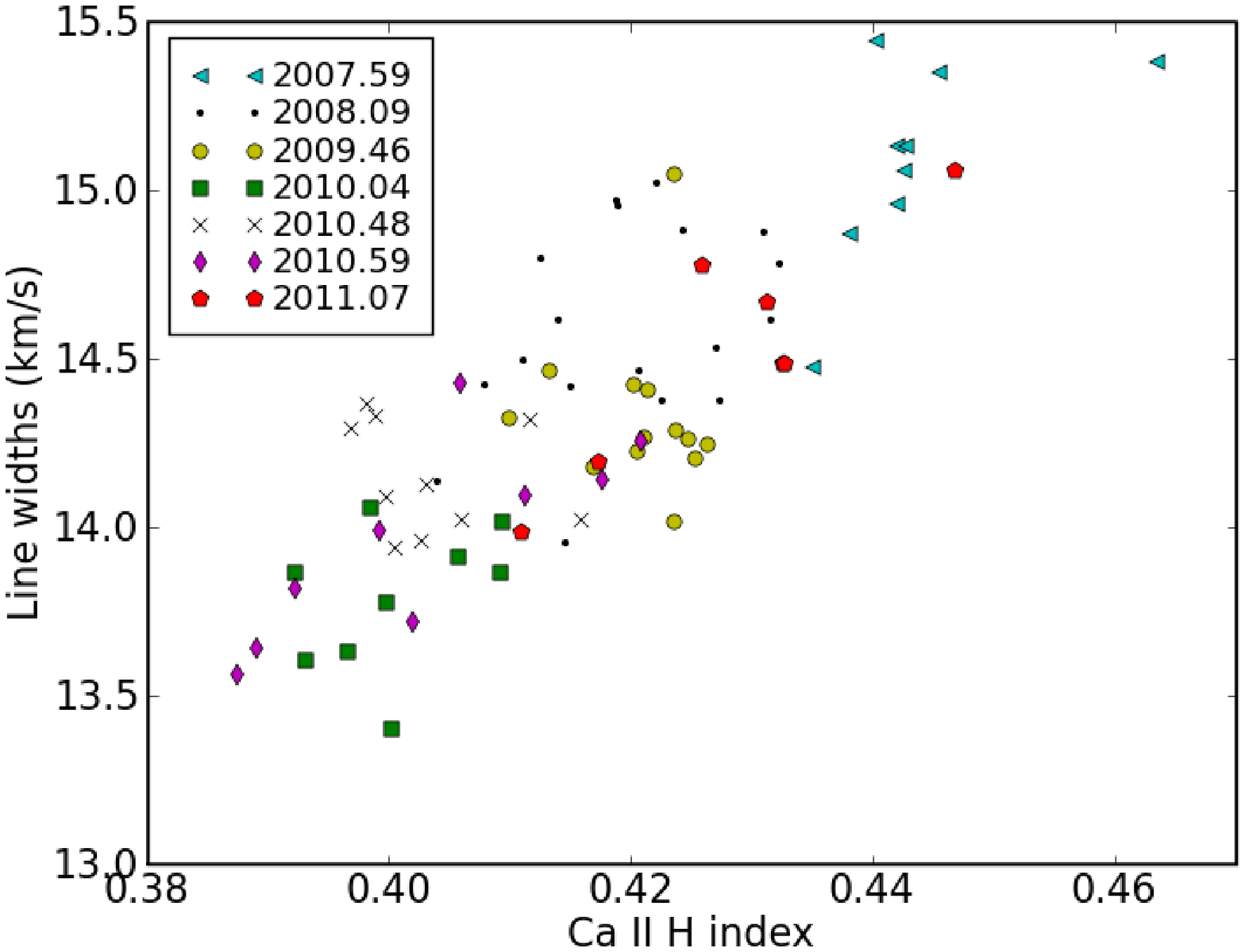}
\caption{Correlation between $N_{CaII H}$ and $N_{H_{\alpha}}$ (top), and between the widths of the FeI@846.84 line and $N_{CaII H}$ (bottom).}
\label{fig:correlation}
\end{figure}

Most conclusions drawn for the $N_{CaII H}$ index are also valid for
$N_{H\alpha}$,  given the correlation of the two indices illustrated in Fig. \ref{fig:correlation}. Again, rotational variability is observed for most epochs, as well as a year-to-year trend. The changes seen between the observing epochs are higher than the rotationally induced variations.

\subsection{Radial velocities and profile bisectors}
\label{sec:radial}

Even for the low Doppler broadening of spectral lines suffered by \xib\ (\vsin$=3$\kms), surface inhomogeneities (cool spots or plages) can be expected to modify the shapes of the Stokes I LSD profiles. During its transit across the visible hemisphere of the star, a cool spot changes the depth of one wing of the profile when it gets closer to the stellar limb (so that its spectral counterpart gains significant Doppler shift). This effect may be detected as a change in the radial velocity of the line centroid, or as a modification of the profile bisector. The typical profile asymmetry produced by convective flows \citep{toner88} may also be locally modified by strong magnetic fields, resulting in a temporary change of the profile bisector. In this section, we propose to investigate these effects for \xib.

Radial velocities (RV hereafter) were calculated from our Stokes I LSD profiles by applying a Gaussian fit to the line profiles and using the centroid of the resulting Gaussian function as a radial velocity estimate. Using a (symmetric) Gaussian fit may be thought inadapted for asymmetric Stokes I LSD profiles observed for \xib. We note, however, that the level of asymmetry is generally low compared to radial velocity fluctuations derived below (Table \ref{tab:moy_ind}). Detailed measurements can be found in Table \ref{tab:tout} and phase-folded results are shown in Figs. \ref{fig:traceurs1} and \ref{fig:traceurs2}. The typical error for a single measure of RV, using NARVAL in polarimetric mode, was estimated by \citet{moutou07} and is of the order of 0.015-0.030 \kms. Several data sets display rotationally modulated variations in RV, with amplitudes sometimes in excess of 0.1 \kms\ (see for instance 2009.46). Most of the time, this activity proxy is acting differently from other tracers. 

\begin{figure*}
\centering
\includegraphics[scale = 0.35]{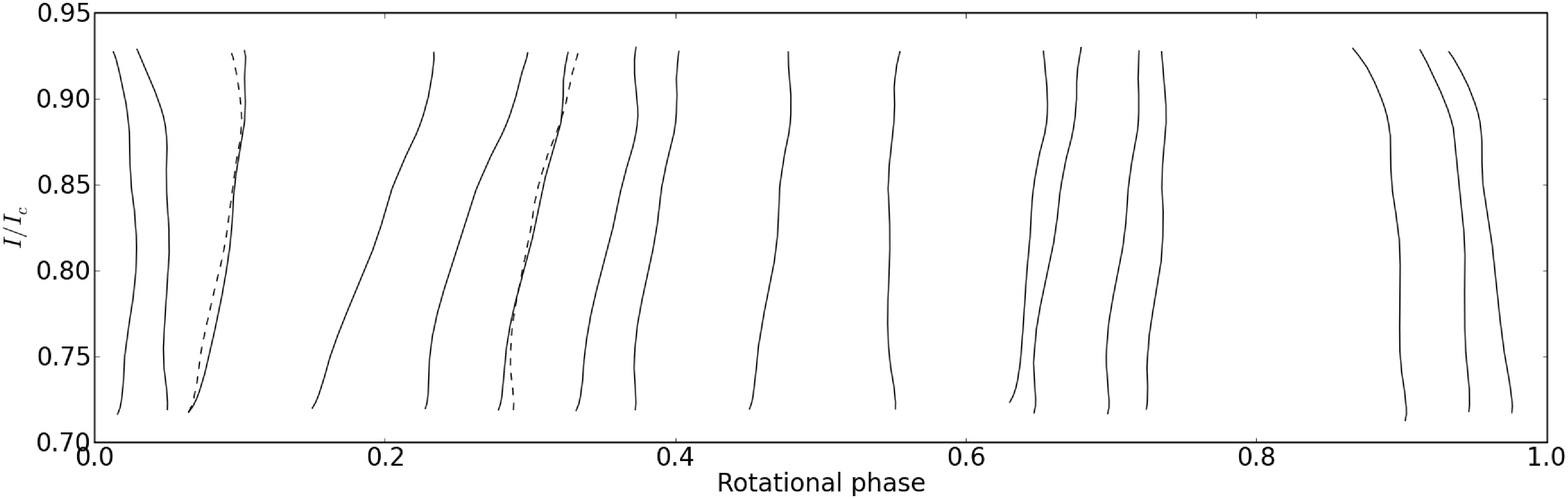}
\caption{Bisectors of Stokes I LSD profiles as a function of the rotational phase for 2008.09. Dashes represent observations taken a few minutes apart from the overplotted solid-line bisector.}
\label{fig:biss_phase}
\end{figure*}

We also used the Stokes I LSD profiles to construct profile bisectors. After adjusting the sampling of the intensity line profile with a cubic spline procedure, we computed the bisector using a method similar to \citet*{toner88}. The phase-dependence of the bisector shape is plotted in Figure \ref{fig:biss_phase} for the observations of 2008.09. We note that the bisectors never display the typical redshift near the continuum that they exhibited most of the time in previous studies of \citet*{toner88} and \citet{petit05}. This shape difference is due to the spectral range (covering the whole visible domain) used to derive the LSD profiles. Using various sub-sets of our line-list to compute LSD profiles in narrower spectral regions, we note that the shape reported by \citet*{toner88} is typical of the red- and near-infrared domains, while a mirror shape (with a blueshift close to the continuum) is observed for the bluest parts of the NARVAL spectral range. 

We note that strong and weak individual spectral lines (which are expected to have different bisector shapes) are mixed together in the computation of a LSD profile. If the resulting bisector shape is still a tracer of photospheric convective flows \citep{gray80, gray81, gray82, dravins87}, any direct comparison with individual line bisectors should be taken with caution. When adopting the exact line list of \citet{toner88} to compute LSD profiles (eight lines close to 625 nm), we indeed derived bisectors similar to theirs, but here again the comparison is of limited relevance, since it might be affected by the difference in spectral resolution and time of observation.  

To quantify the changes in the bisector shape, we calculated the velocity span associated to each bisector, as defined by  \citet*{toner88}. Here, we take the difference in radial velocity between a point near the top of the line profile ($I/I_{c}=0.95$) and a point near the bottom (the $I/I_{c}$ minimum value plus 0.01). Uncertainties in the bisector span were derived using the approach of \citet{povich01}, yielding error bars of 20-30 \ms. Repeated measurements within a few minutes, obtained at two occurences in 2008.09, provided us with variations of the velocity span within the calculated uncertainties. The bisector fluctuations are shown in Figures \ref{fig:traceurs1} and \ref{fig:traceurs2}, with numerical quantities listed in Tab. \ref{tab:tout}. Variations with the rotational phase are visible for most epochs of observation, and sometimes follow a complex pattern (e.g. 2008.09). Most of the time, no obvious correlation can be found with other magnetic or activity proxies. In contrast to most other tracers, a long-term trend of the bisector span is not detected above its typical dispersion level (Fig. \ref{fig:val_moy}).

\section{Discussion and conclusions}

We have used spectropolarimetric observations to derive a time-series of magnetic field maps and activity proxies covering almost four years in the life of \xib. In doing so we combined observational approaches that are often investigated separately.

\subsection{Magnetic topology}

In spite of its rapid rotation rate, the spectral lines of \xib\ are not significantly affected by Doppler broadening. The low \vsin\ of the star is due to the combined effect of a low inclination angle and a relatively small radius. When applied to stars with low rotational broadening, ZDI selectively reconstructs only the largest spatial scales of the surface magnetic field. The reconstructed magnetic maps of \xib\ show a surface-averaged field strength of the order of 30-100~G. These values are significantly lower than the previous estimate of \citet{petit05}, derived from observations taken during the summer of 2003, both for the poloidal and toroidal field components. This difference is possibly the consequence of a long-term magnetic trend, however, we emphasize that the instrumental setup and modelling methods are different in the two studies \citep[direct modelling in][inverse tomography in the present study]{petit05}, which may account for part of the apparent decrease. Both methods agree on the existence of a strong toroidal field component on \xib, which is dominating the energy budget in the results of \citet{petit05}, and also in most of the ZDI magnetic field reconstructions in the present study, except in 2008.09 and 2010.48 (for which the poloidal field component dominates the surface magnetic energy).

The properties of the large-scale field of \xib\ can also be compared to that of other ZDI studies of solar-type dwarfs, as long as the selected stars also exhibit a low \vsin, so as to ensure that all stars used for the comparison are affected by a same low-pass spatial filtering of their magnetic geometry. As far as mass is concerned, the closest available neighbour is HD~189733 \citep{fares10}. The average surface field strength recovered for this star is lower than that derived for \xib, which is likely a consequence of the slower rotation rate of HD~189733 (with a rotation period of 12~d). In spite of this difference, both stars have a significant fraction of their surface magnetic energy reconstructed in a toroidal field component.

A small number of main-sequence stars of higher mass can also serve as comparisons. The average field strength of \xib\ is consistent with the large-scale field strength measured for 1~\msun\ stars with rotation periods slower than about 10~d \citep{petit08}. The prominent toroidal component of the large-scale magnetic geometry, observed for each year except for 2008.09 and 2010.48, agrees with the assertion, derived from both observations and numerical simulations of stellar dynamos, that surface magnetic geometries become predominantly toroidal for solar-type dwarfs with rotation periods as short as a few days, whereas a mainly poloidal surface field is a common feature of slow rotators \citep{petit08,brown10}.

\subsection{Magnetic variability}

If \xib\ seems to obey the above-mentioned trends, the temporal fluctuations of its magnetic properties demonstrate that magnetic variability will induce some scatter around any average behaviour, because the global field strength, the fraction of energy in the toroidal field, and the field axisymmetry all vary with time. While relative fluctuations observed in the chromospheric emission are limited to about 20\% \citep[although][suggest that longer-term variations of up to 60\% can be observed]{baliunas95}, simultaneous variations by a factor of about 2 are observed in the mean large-scale field strength, and reach a factor of about 5 in the fraction of energy reconstructed in the axisymmetric field component and a factor of 3 in the fraction of energy stored as a toroidal component. Assuming that these sharp variations are too important to simply reflect intrinsic limitations of our magnetic model (Sect. \ref{sec:maps}), they suggest that the large-scale field structures seen in polarimetry are more sensitive to dynamo fluctuations than the smaller-scale structures that contribute to the chromospheric flux. The difference in behaviour between the activity tracers and the large-scale field is also visible on the shorter term, e.g., between the close-by epochs of 2010.48 and 2010.59, when a fast evolution of the global magnetic topology \citep[similar in its short timescale to previous observations by][]{petit05} has no clear counterpart in other activity proxies.

Several possibilities can be proposed to account for the low agreement between the Stokes V and Stokes I activity indicators. Firstly, Zeeman broadening and chromospheric emission are expected to trace a wider range of spatial scales than the polarized Zeeman signatures, which are missing a significant fraction of the total magnetic flux of the star. Secondly, the different limb visibility of the various proxies is expected to result in different temporal evolutions. Thirdly, the flux in chromospheric indicators tends to saturate with increasing magnetic flux \citep{schrijver89,loukitcheva09}, so that strong magnetic field variations may result in more gentle Ca~II variations.

The complex long-term magnetic variations of \xib, previously noted by \citet{baliunas95}, are reminiscent of the behaviour of high-activity, very rapidly-rotating dwarfs \citep{donati03dwarf,jeffers11,marsden11}. There is obvious magnetic field evolution over the four years of observations, but it does not take the simple form of fast global polarity switches observed on other rapidly-rotating sun-like stars \citep{fares09, petit09, morgenthaler11}. Instead, the main variability observed is a drop of the average field strength from 2008 to early 2010, encompassed by epochs associated with stronger field. This evolution is also observed in most of the activity proxies. We note that the toroidal field component is relatively weaker whenever the global field strength is close to its minimum value, and observe a more axisymmetric field structure in high-activity states. The magnetic topologies of 2007.59 and 2011.07 are very similar, suggesting that the repeated observation of this magnetic pattern may be related to a type of magnetic cycle that does not imply a global polarity switch.  Based on our data sets, the large-scale magnetic variability of \xib\ may seem chaotic, however, we stress that not all relevant temporal scales are explored here, so that a longer monitoring or a denser temporal sampling may help reveal other patterns.

\subsection{Differential rotation}

Some of the short-term changes in the magnetic topology can be taken into account in our model, assuming that they obey a simple differential rotation law. The measurement of the rotational shear was not conclusive at every epoch, presumably because of the fast emergence/disappearance of magnetic spots, which affects the tracking accuracy of the magnetic tracers used to determine the shear \citep[a source of error already mentioned by][]{petit02}. Whenever measurable, the values of the \drot\ parameters are consistent with a strong surface shear, with \dom\ from five to ten times the solar value. As mentioned in Sec. \ref{sec:maps}, part of this observed scatter is likely generated by fast changes in the surface topology, but this random effect is not expected to result in systematic biases of \dom\ measurements. The \dom\ values are far above some measurements obtained for very rapidly rotating dwarfs in this temperature domain \citep[see][who report a drop of \drot\ for stars cooler than the Sun]{barnes05}. It is, however, reminiscent of the strong shear observed on the T Tauri star v2247 Oph, in spite of a spectral type as late as M2 \citep{donati10}. An increase in \dom\ with rotation is also consistent with other observational studies \citep{donahue96,saar09} and with the models of \citet{brown08} and \citet{ballot07}. 

The high latitudinal shear of \xib\ also suggests that in spite of the high activity level of the star, Maxwell stresses are still inefficient at forcing solid-body rotation as proposed, e.g., for active M dwarfs \citep{morin08,browning08}, but also for very rapidly rotating Sun-like stars \citep{henry96,reiners03}. If such an extreme state is not reached by \xib, the sharp apparent fluctuations of the shear level, if real, may indicate that a feedback of the magnetic field onto the large-scale flows is probably active in the convective layers and is affecting the \drot, as suggested for other high-activity dwarfs \citep*{donati03} and from dynamo models of \citet{brun05}.

\begin{acknowledgements}
This research made use of the POLLUX database (http://pollux.graal.univ-monpt2.fr) operated at LUPM (Universit\'e Montpellier II - CNRS, France, with support of the PNPS and INSU). We are grateful to the staff of TBL for their efficient help during the many nights dedicated to this observing project. We are grateful to an anonymous referee for a number of constructive comments that helped to clarify this article. This long-term magnetic monitoring was undertaken in the framework of the Bcool project.
\end{acknowledgements}

\bibliography{ksibooa}

\onltab{1}{
\setcounter{table}{2}
\begin{table}
\caption{Journal of observations for 2007.59, 2008.09, and 2009.46. From left to right, we list the year of observation, the Julian date, the error-bar in Stokes V LSD profiles, and the phase of the rotational cycle at which the observation was made, taking the same rotation period and phase origin as \citet{petit05}.}
\begin{tabular}{cccccc}
\hline
Year & Julian date & $\sigma_{\rm LSD}$ & rot. phase \\
 & (2,450,000+) & $10^{-5}I_{c}$ & & \\
\hline
2007.59 & 4308.35 & 5.4742 & 0.8726 \\
& 4310.35 & 1.7360 & 0.1835 \\
& 4311.35 & 2.1059 & 0.3387 \\
& 4312.34 & 2.7435 & 0.4928 \\
& 4313.34 & 3.8929 & 0.6485 \\
& 4315.34 & 2.8654 & 0.9586 \\
& 4316.34 & 2.7608 & 0.1152 \\
& 4322.34 & 3.2082 & 0.0473 \\
& 4323.34 & 3.7479 & 0.2030 \\
\hline
2008.09 & 4484.72 & 5.9851  & 0.3017 \\
& 4484.77 & 6.3988 & 0.3092 \\
& 4485.77 & 5.2155 & 0.4648 \\
& 4488.76 & 3.2560 & 0.9291 \\
& 4489.74 & 3.4732 & 0.0827 \\
& 4489.75 & 3.5767 & 0.0840 \\
& 4491.70 & 4.0728 & 0.3865 \\
& 4492.75 & 3.3416 & 0.5499 \\
& 4493.76 & 2.7275 & 0.7075 \\
& 4495.77 & 3.4875 & 0.0206 \\
& 4499.77 & 12.7455  & 0.6427 \\
& 4501.77 & 3.5052  & 0.9538 \\
& 4503.76 & 3.1664 & 0.2626 \\
& 4506.77 & 2.9219  & 0.7304 \\
& 4507.76 & 1.3428  & 0.8843 \\
& 4508.76 & 4.5386  & 0.0400 \\
& 4509.73 & 7.8197  & 0.1914 \\
& 4510.77 & 4.1865  & 0.3524 \\
& 4512.76 & 3.1274  & 0.6627 \\
 \hline
2009.46 & 4980.38 & 4.0385 & 0.3865 \\
 & 4984.45 & 12.1589  & 0.0192 \\
 & 4985.43 & 3.7381 & 0.1722 \\
 & 4994.47 & 5.3269 & 0.5786 \\
 & 4995.46 & 4.3773 & 0.7318 \\
 & 5001.44 & 14.4977 & 0.6617 \\
 & 5002.36 & 10.5647 & 0.8056 \\
 & 5006.56 & 3.2475 & 0.4579 \\
 & 5010.46 & 5.4779 & 0.0654 \\
 & 5011.44 & 4.6241 & 0.2175 \\
 & 5013.55 & 4.3254 & 0.5450 \\
& 5017.38 & 6.9613 & 0.1415 \\
 & 5018.40 & 3.1676 & 0.2992 \\
\hline
\end{tabular}
\label{tab:obs1}
\end{table}
}

\onltab{2}{
\setcounter{table}{3}
\begin{table}
\caption{Same as \ref{tab:obs1} for 2010.04, 2010.48, 2010.59,
  and 2011.07.}
\begin{tabular}{cccccc}
\hline
Year & Julian date & $\sigma_{\rm LSD}$ & rot. phase \\
 & (2,450,000+) & $10^{-5}I_{c}$ & & \\
\hline
2010.04 & 5180.78 & 8.7507 & 0.5527 \\
& 5181.78 & 3.2145 & 0.7082 \\
& 5202.71 & 4.7321 & 0.9641 \\
& 5215.65 & 4.5171 & 0.9763 \\
& 5222.71 & 3.1795 & 0.0743 \\
& 5224.69 & 3.2308 & 0.3823 \\
& 5240.62 & 4.6358 & 0.8593 \\
& 5241.66 & 7.1186 & 0.0211 \\
& 5242.63 & 3.7747 & 0.1718 \\
\hline
2010.48  & 5354.42 & 3.9811 & 0.5580 \\
& 5370.45 & 4.8856 & 0.0514 \\
& 5379.41 & 3.8653 & 0.4445 \\
& 5382.37 & 3.0234 & 0.9054 \\
& 5383.37 & 2.9261 & 0.0600 \\
& 5384.37 & 2.6600 & 0.2154 \\
& 5388.36 & 2.9740 & 0.8374 \\
& 5390.37 & 5.0528 & 0.1487 \\
& 5391.40 & 2.9108 & 0.3093 \\
& 5392.40 & 2.3126 & 0.4650 \\
\hline
2010.59 & 5403.41 & 3.0789 & 0.1766 \\
& 5412.36 & 3.4796 & 0.5697 \\
& 5414.35 & 2.8979 & 0.8791 \\
& 5415.35 & 2.9610 & 0.0335 \\
& 5416.36 & 4.4882 & 0.1919 \\
& 5417.35 & 3.6313 & 0.3448 \\
& 5418.38 & 3.6560 & 0.5047 \\
& 5421.36 & 3.2580 & 0.9690 \\
& 5427.34 & 2.9531 & 0.8985 \\
\hline
2011.07 & 5578.71 & 3.6620 & 0.4409 \\
& 5584.76 & 3.3521 & 0.3814 \\
& 5586.75 & 3.1925 & 0.6901 \\
& 5587.75 & 3.1746 & 0.8459 \\
& 5588.76 & 2.9784 & 0.0028 \\
& 5593.75 & 4.0095 & 0.7787 \\
& 5596.71 & 4.1288 & 0.2387 \\
\hline
\end{tabular}
\label{tab:obs2}
\end{table}
}

\onltab{3}{
\setcounter{table}{4}
\begin{table*}
\caption{Activity tracers of $\xi$ Bootis A derived from the Stokes I
  profiles for each observation for 2007.59, 2008.09, and 2009.46.}
\begin{center}
\begin{tabular}{cccccccc}
\hline
Frac. year & Julian date & $B_l$ & Rad. vel.  & Velocity spans & Line widths & $N_{CaII H}$ & $N_{H_{\alpha}}$ \\
 & & (G) & (km.s$^{-1}$) & (km.s$^{-1}$) & (km.s$^{-1}$) & & \\
\hline 
 2007.59 & 2454315.34 & $22.0 \pm 0.8$ & $1.86$ & $0.02 \pm 0.02$ & $15.059 \pm 0.041$ & $0.4425\pm0.0005$ & $0.3577\pm0.0002$ \\
 & 2454316.34 & $8.5 \pm 0.9$ & $1.86$ & $0.02 \pm 0.02$ & $15.135 \pm 0.038$ & $0.442\pm0.0005$ & $0.3584\pm0.0002$ \\
 & 2454322.34 & $11.7 \pm 0.9$ & $1.84$ & $0.03 \pm 0.02$ & $15.133 \pm 0.041$ & $0.4427\pm0.0005$ & $0.3614\pm0.0002$ \\
 & 2454323.34 & $-0.6 \pm 1.1$ & $1.87$ & $0.03 \pm 0.03$ & $15.385 \pm 0.037$ & $0.4635\pm0.0006$ & $0.3631\pm0.0002$ \\
 & 2454308.35 & $18.2 \pm 1.7$ & $1.88$ & $0.06 \pm 0.02$ & $15.444 \pm 0.007$ & $0.4401\pm0.0011$ & $0.3571\pm0.0004$ \\
 & 2454310.35 & $7.3 \pm 0.5$ & $1.88$ & $0.03 \pm 0.02$ & $14.96 \pm 0.032$ & $0.4419\pm0.0003$ & $0.358\pm0.0001$ \\
 & 2454311.35 & $1.4 \pm 0.6$ & $1.85$ & $0.04 \pm 0.02$ & $15.352 \pm 0.037$ & $0.4454\pm0.0003$ & $0.3585\pm0.0001$ \\
 & 2454312.34 & $2.8 \pm 0.8$ & $1.82$ & $0.07 \pm 0.02$ & $14.875 \pm 0.032$ & $0.438\pm0.0004$ & $0.3582\pm0.0002$ \\
 & 2454313.34 & $6.9 \pm 1.1$ & $1.84$ & $0.04 \pm 0.02$ & $14.477 \pm 0.037$ & $0.435\pm0.0006$ & $0.3566\pm0.0002$ \\
\hline
 2008.09 & 2454499.77 & $7.4 \pm 3.6$ & $1.67$ & $0.05 \pm 0.02$ & $14.799 \pm 0.023$ & $0.4124\pm0.0018$ & $0.3542\pm0.0006$ \\
 & 2454501.77 & $9.8 \pm 1.0$ & $1.8$ & $0.07 \pm 0.03$ & $14.616 \pm 0.041$ & $0.4138\pm0.0006$ & $0.3562\pm0.0002$ \\
 & 2454503.76 & $4.5 \pm 0.9$ & $1.71$ & $0.08 \pm 0.02$ & $14.876 \pm 0.034$ & $0.4309\pm0.0005$ & $0.3557\pm0.0002$ \\
 & 2454506.77 & $3.1 \pm 0.9$ & $1.73$ & $0.02 \pm 0.02$ & $14.137 \pm 0.036$ & $0.4039\pm0.0005$ & $0.3535\pm0.0002$ \\
 & 2454507.76 & $5.9 \pm 0.4$ & $1.75$ & $0.07 \pm 0.02$ & $14.467 \pm 0.036$ & $0.4206\pm0.0002$ & $0.354\pm0.0001$ \\
 & 2454508.76 & $0.8 \pm 1.3$ & $1.72$ & $0.03 \pm 0.02$ & $14.783 \pm 0.037$ & $0.4322\pm0.0007$ & $0.3583\pm0.0003$ \\
 & 2454509.73 & $6.7 \pm 2.0$ & $1.73$ & $0.09 \pm 0.02$ & $14.97 \pm 0.022$ & $0.4187\pm0.0012$ & $0.357\pm0.0005$ \\
 & 2454510.77 & $4.6 \pm 1.1$ & $1.68$ & $0.05 \pm 0.02$ & $14.533 \pm 0.036$ & $0.427\pm0.0007$ & $0.3549\pm0.0003$ \\
 & 2454512.76 & $7.1 \pm 0.8$ & $1.65$ & $0.04 \pm 0.02$ & $13.956 \pm 0.024$ & $0.4145\pm0.0005$ & $0.3526\pm0.0002$ \\
 & 2454484.72 & $3.9 \pm 2.1$ & $1.73$ & $0.06 \pm 0.02$ & $14.885 \pm 0.038$ & $0.4242\pm0.0012$ & $0.3565\pm0.0004$ \\
 & 2454484.77 & $6.4 \pm 2.1$ & $1.73$ & $0.05 \pm 0.02$ & $14.379 \pm 0.031$ & $0.4224\pm0.0012$ & $0.3573\pm0.0004$ \\
 & 2454485.77 & $3.9 \pm 1.5$ & $1.72$ & $0.04 \pm 0.02$ & $14.412 \pm 0.038$ & $0.4212\pm0.0009$ & $0.3541\pm0.0003$ \\
 & 2454488.76 & $1.8 \pm 0.9$ & $1.84$ & $0.06 \pm 0.03$ & $14.417 \pm 0.037$ & $0.4149\pm0.0005$ & $0.353\pm0.0002$ \\
 & 2454489.75 & $5.2 \pm 1.0$ & $1.82$ & $0.05 \pm 0.02$ & $15.025 \pm 0.038$ & $0.4221\pm0.0005$ & $0.3563\pm0.0002$ \\
 & 2454489.74 & $4.7 \pm 1.0$ & $1.82$ & $0.04 \pm 0.02$ & $14.958 \pm 0.032$ & $0.4188\pm0.0005$ & $0.3562\pm0.0002$ \\
 & 2454491.7 & $-0.4 \pm 1.2$ & $1.75$ & $0.03 \pm 0.02$ & $14.38 \pm 0.036$ & $0.4273\pm0.0007$ & $0.3555\pm0.0003$ \\
 & 2454492.75 & $-2.3 \pm 1.0$ & $1.73$ & $0.01 \pm 0.02$ & $14.495 \pm 0.037$ & $0.411\pm0.0005$ & $0.3538\pm0.0002$ \\
 & 2454493.76 & $3.7 \pm 0.8$ & $1.76$ & $0.03 \pm 0.02$ & $14.426 \pm 0.042$ & $0.4079\pm0.0004$ & $0.354\pm0.0002$ \\
 & 2454495.77 & $10.5 \pm 1.0$ & $1.78$ & $0.04 \pm 0.02$ & $14.619 \pm 0.037$ & $0.4314\pm0.0006$ & $0.3562\pm0.0002$ \\
\hline
 2009.46 & 2454984.45 & $1.9 \pm 3.5$ & $1.81$ & $0.03 \pm 0.02$ & $14.324 \pm 0.029$ & $0.4098\pm0.0022$ & $0.3537\pm0.0007$ \\
 & 2454985.43 & $3.3 \pm 1.1$ & $1.82$ & $0.04 \pm 0.02$ & $14.228 \pm 0.023$ & $0.4205\pm0.0006$ & $0.3544\pm0.0002$ \\
 & 2455017.38 & $10.4 \pm 1.8$ & $1.88$ & $0.03 \pm 0.02$ & $14.423 \pm 0.038$ & $0.4201\pm0.0011$ & $0.355\pm0.0004$ \\
 & 2455018.4 & $7.5 \pm 0.9$ & $1.89$ & $0.03 \pm 0.03$ & $14.206 \pm 0.018$ & $0.4252\pm0.0005$ & $0.3526\pm0.0002$ \\
 & 2454994.47 & $16.7 \pm 1.6$ & $1.9$ & $0.06 \pm 0.03$ & $14.177 \pm 0.017$ & $0.4168\pm0.001$ & $0.3515\pm0.0003$ \\
 & 2454995.46 & $16.1 \pm 1.4$ & $1.9$ & $0.05 \pm 0.02$ & $14.467 \pm 0.022$ & $0.4132\pm0.0008$ & $0.3537\pm0.0003$ \\
 & 2455001.44 & $22.1 \pm 4.0$ & $1.89$ & $0.05 \pm 0.02$ & $14.017 \pm 0.026$ & $0.4235\pm0.0036$ & $0.3521\pm0.0008$ \\
 & 2455002.36 & $9.9 \pm 2.9$ & $1.88$ & $0.03 \pm 0.02$ & $15.047 \pm 0.036$ & $0.4235\pm0.0015$ & $0.355\pm0.0005$ \\
 & 2455006.56 & $7.7 \pm 0.9$ & $1.89$ & $0.01 \pm 0.02$ & $14.409 \pm 0.036$ & $0.4212\pm0.0006$ & $0.3525\pm0.0002$ \\
 & 2455010.46 & $4.8 \pm 1.4$ & $1.91$ & $0.03 \pm 0.02$ & $14.249 \pm 0.036$ & $0.4263\pm0.0009$ & $0.3541\pm0.0003$ \\
 & 2455011.44 & $3.3 \pm 1.3$ & $1.9$ & $0.03 \pm 0.02$ & $14.271 \pm 0.037$ & $0.421\pm0.0008$ & $0.3529\pm0.0003$ \\
 & 2454980.38 & $6.5 \pm 1.2$ & $1.85$ & $0.04 \pm 0.02$ & $14.287 \pm 0.033$ & $0.4236\pm0.0007$ & $0.3534\pm0.0003$ \\
 & 2455013.55 & $-1.8 \pm 1.3$ & $1.88$ & $0.03 \pm 0.02$ & $14.265 \pm 0.037$ & $0.4247\pm0.0009$ & $0.3512\pm0.0003$ \\
\hline
\end{tabular}
\end{center}
\label{tab:tout}
\end{table*}
}

\onltab{3}{
\setcounter{table}{5}
\begin{table*}
\caption{Same as \ref{tab:tout} for 2010.04, 2010.48, 2010.59, and 2011.07.}
\begin{center}
\begin{tabular}{cccccccc}
\hline
Frac. year & Julian date & $B_l$ & Rad. vel.  & Velocity spans & Line widths & $N_{CaII H}$ & $N_{H_{\alpha}}$ \\
 & & (G) & (km.s$^{-1}$) & (km.s$^{-1}$) & (km.s$^{-1}$) & & \\
\hline
 2010.04 & 2455202.71 & $-1.1 \pm 1.4$ & $1.77$ & $0.02 \pm 0.02$ & $14.016 \pm 0.038$ & $0.4093\pm0.0008$ & $0.353\pm0.0003$ \\
 & 2455240.62 & $13.7 \pm 1.4$ & $1.78$ & $0.02 \pm 0.02$ & $13.607 \pm 0.034$ & $0.393\pm0.0008$ & $0.3508\pm0.0003$ \\
 & 2455241.66 & $10.3 \pm 1.8$ & $1.74$ & $0.02 \pm 0.02$ & $13.633 \pm 0.033$ & $0.3965\pm0.0013$ & $0.3515\pm0.0004$ \\
 & 2455180.78 & $7.3 \pm 2.4$ & $1.78$ & $0.05 \pm 0.03$ & $14.057 \pm 0.037$ & $0.3983\pm0.0019$ & $0.3501\pm0.0005$ \\
 & 2455242.63 & $-2.5 \pm 1.1$ & $1.79$ & $0.02 \pm 0.02$ & $13.913 \pm 0.017$ & $0.4056\pm0.0007$ & $0.3524\pm0.0002$ \\
 & 2455181.78 & $-0.1 \pm 0.8$ & $1.68$ & $0.04 \pm 0.02$ & $13.87 \pm 0.037$ & $0.4091\pm0.0005$ & $0.3523\pm0.0002$ \\
 & 2455215.65 & $3.7 \pm 1.3$ & $1.83$ & $0.04 \pm 0.03$ & $13.779 \pm 0.038$ & $0.3997\pm0.001$ & $0.351\pm0.0003$ \\
 & 2455222.71 & $5.6 \pm 0.9$ & $1.8$ & $0.02 \pm 0.02$ & $13.87 \pm 0.037$ & $0.3921\pm0.0005$ & $0.3502\pm0.0002$ \\
 & 2455224.69 & $-0.4 \pm 1.0$ & $1.84$ & $0.03 \pm 0.02$ & $13.403 \pm 0.018$ & $0.4001\pm0.0006$ & $0.35\pm0.0002$ \\
\hline
 2010.48 & 2455379.41 & $12.2 \pm 1.1$ & $1.89$ & $0.07 \pm 0.02$ & $13.941 \pm 0.038$ & $0.4003\pm0.0007$ & $0.3489\pm0.0002$ \\
 & 2455412.36 & $10.5 \pm 0.8$ & $1.93$ & $0.04 \pm 0.02$ & $14.318 \pm 0.037$ & $0.4115\pm0.0005$ & $0.3523\pm0.0002$ \\
 & 2455382.37 & $10.7 \pm 0.9$ & $1.94$ & $0.01 \pm 0.02$ & $14.093 \pm 0.04$ & $0.3997\pm0.0005$ & $0.3511\pm0.0002$ \\
 & 2455414.35 & $4.3 \pm 0.8$ & $1.93$ & $0.04 \pm 0.02$ & $14.331 \pm 0.026$ & $0.3988\pm0.0004$ & $0.3492\pm0.0002$ \\
 & 2455383.37 & $11.2 \pm 1.1$ & $1.89$ & $0.03 \pm 0.02$ & $14.128 \pm 0.008$ & $0.403\pm0.0006$ & $0.3514\pm0.0002$ \\
 & 2455415.35 & $5.2 \pm 0.8$ & $1.95$ & $0.07 \pm 0.03$ & $14.024 \pm 0.04$ & $0.4157\pm0.0005$ & $0.3498\pm0.0002$ \\
 & 2455384.37 & $10.5 \pm 1.6$ & $1.98$ & $0.03 \pm 0.03$ & $13.962 \pm 0.036$ & $0.4025\pm0.001$ & $0.3515\pm0.0002$ \\
 & 2455354.42 & $3.3 \pm 0.9$ & $1.95$ & $0.02 \pm 0.02$ & $14.292 \pm 0.038$ & $0.3967\pm0.0005$ & $0.3514\pm0.0002$ \\
 & 2455416.36 & $3.5 \pm 0.7$ & $1.92$ & $0.04 \pm 0.02$ & $14.023 \pm 0.037$ & $0.4059\pm0.0004$ & $0.3512\pm0.0003$ \\
 & 2455417.35 & $14.4 \pm 1.4$ & $1.89$ & $0.04 \pm 0.02$ & $14.365 \pm 0.041$ & $0.398\pm0.0009$ & $0.3502\pm0.0002$ \\
\hline
 2010.59 & 2455418.38 & $6.9 \pm 1.0$ & $1.92$ & $0.03 \pm 0.02$ & $14.095 \pm 0.04$ & $0.4111\pm0.0006$ & $0.3522\pm0.0002$ \\
 & 2455388.36 & $12.3 \pm 0.8$ & $1.91$ & $0.02 \pm 0.02$ & $13.566 \pm 0.037$ & $0.3873\pm0.0004$ & $0.3516\pm0.0002$ \\
 & 2455421.36 & $17.4 \pm 0.8$ & $1.97$ & $0.05 \pm 0.02$ & $13.82 \pm 0.039$ & $0.3921\pm0.0005$ & $0.3496\pm0.0002$ \\
 & 2455390.37 & $17.6 \pm 1.3$ & $1.97$ & $0.05 \pm 0.03$ & $14.429 \pm 0.038$ & $0.4058\pm0.0008$ & $0.3508\pm0.0003$ \\
 & 2455391.4 & $7.8 \pm 1.0$ & $1.95$ & $0.05 \pm 0.02$ & $14.143 \pm 0.037$ & $0.4175\pm0.0006$ & $0.3514\pm0.0002$ \\
 & 2455392.4 & $4.3 \pm 1.0$ & $1.92$ & $0.06 \pm 0.02$ & $14.256 \pm 0.025$ & $0.4207\pm0.0006$ & $0.3502\pm0.0002$ \\
 & 2455427.34 & $12.8 \pm 1.0$ & $1.92$ & $0.05 \pm 0.02$ & $13.641 \pm 0.039$ & $0.3889\pm0.0006$ & $0.3469\pm0.0002$ \\
 & 2455370.45 & $10.8 \pm 0.9$ & $1.93$ & $0.04 \pm 0.02$ & $13.723 \pm 0.031$ & $0.4018\pm0.0005$ & $0.3537\pm0.0003$ \\
 & 2455403.41 & $10.7 \pm 0.9$ & $1.95$ & $0.08 \pm 0.02$ & $13.992 \pm 0.033$ & $0.399\pm0.0005$ & $0.3514\pm0.0002$ \\
\hline
 2011.07 & 2455596.71 & $7.3 \pm 1.1$ & $1.84$ & $0.02 \pm 0.02$ & $15.061 \pm 0.037$ & $0.4467\pm0.0007$ & $0.3615\pm0.0003$ \\
 & 2455578.71 & $8.0 \pm 0.9$ & $1.85$ & $0.07 \pm 0.03$ & $14.777 \pm 0.027$ & $0.4258\pm0.0006$ & $0.3584\pm0.0002$ \\
 & 2455584.76 & $6.9 \pm 0.9$ & $1.81$ & $0.05 \pm 0.02$ & $14.489 \pm 0.025$ & $0.4325\pm0.0005$ & $0.3585\pm0.0002$ \\
 & 2455586.75 & $6.9 \pm 0.9$ & $1.86$ & $0.03 \pm 0.02$ & $14.489 \pm 0.025$ & $0.4326\pm0.0005$ & $0.3541\pm0.0002$ \\
 & 2455587.75 & $6.8 \pm 0.8$ & $1.85$ & $0.03 \pm 0.02$ & $14.198 \pm 0.033$ & $0.4172\pm0.0005$ & $0.3558\pm0.0002$ \\
 & 2455588.76 & $15.9 \pm 0.8$ & $1.8$ & $0.04 \pm 0.02$ & $14.667 \pm 0.032$ & $0.4311\pm0.0005$ & $0.3586\pm0.0002$ \\
 & 2455593.75 & $7.3 \pm 1.1$ & $1.82$ & $0.05 \pm 0.03$ & $13.985 \pm 0.035$ & $0.4109\pm0.0007$ & $0.3542\pm0.0002$ \\
\hline
\end{tabular}
\end{center}
\label{tab:tout2}
\end{table*}
}

\end{document}